\documentclass[reprint,aps,prb,amsmath,twocolumn,preprintnumber,superscriptaddress,showpacs,notitlepage,nofootinbib,longbibliography]{revtex4-2}
\usepackage[T1]{fontenc}
\usepackage[latin9]{inputenc}
\makeatletter

\usepackage{mathptmx,newtxtext,newtxmath,xspace}
\usepackage{amsbsy,bm}
\usepackage{float}
\usepackage{graphicx,color,xcolor,epsfig,rotate}
\usepackage{fancyhdr}
\usepackage{comment}
\usepackage{placeins}
\graphicspath{ {images/} }
\usepackage[colorlinks=true, 
            linkcolor=blue, 
            urlcolor=blue,
            citecolor=blue]{hyperref}
\usepackage{soul}
\usepackage{cancel}
\allowdisplaybreaks[2]

\usepackage{diagbox}

\newcommand{\iu}{{i\mkern1mu}}

\makeatother

\usepackage{babel}
\begin{document}
\title{Machine learning assisted derivation of effective low energy models for metallic magnets}
\author{Vikram~Sharma}
\affiliation{Department of Physics and Astronomy, The University of Tennessee,
Knoxville, Tennessee 37996, USA}
\author{Zhentao Wang}
\affiliation{Center for Correlated Matter and Department of Physics, Zhejiang University, Hangzhou 310058, China}
\affiliation{School of Physics and Astronomy, University of Minnesota, Minneapolis, Minnesota 55455, USA}
\affiliation{Department of Physics and Astronomy, The University of Tennessee,
Knoxville, Tennessee 37996, USA}
\author{Cristian~D.~Batista}
\affiliation{Department of Physics and Astronomy, The University of Tennessee,
Knoxville, Tennessee 37996, USA}
\affiliation{Quantum Condensed Matter Division and Shull-Wollan Center, Oak Ridge
National Laboratory, Oak Ridge, Tennessee 37831, USA}
\date{\today}
\begin{abstract}
We consider the problem of extracting an effective low-energy  spin model from a Kondo Lattice Model (KLM) with classical localized moments. The non-analytic dependence of the effective spin-spin interactions on the Kondo exchange $J$ excludes the possibility of using perturbation theory beyond the second order Ruderman-Kittel-Kasuya-Yosida (RKKY) interaction at zero temperature. Here we introduce a Machine Learning (ML) assisted protocol to extract effective two- and four-spin interactions by integrating out the conduction electrons of the original KLM. The resulting effective spin model reproduces the  phase diagram obtained with the original KLM as a function of magnetic field and easy-axis anisotropy and reveals the effective four-spin interactions that are responsible for the field induced skyrmion crystal phase.
Moreover, this minimal spin model enables an efficient computation of static and dynamical properties with a much lower numerical cost relative to the original KLM.  A comparison of the dynamical spin structure factor in the fully polarized phase computed with the effective model and the original KLM reveals a good agreement for the magnon dispersion despite the fact that this information was not included in the training data set. 
\end{abstract}
\maketitle

\section{Introduction}

Lattice models of fermions interacting with classical fields encompass different areas of knowledge, including quantum chemistry, condensed matter, and high-energy physics. 
This broad class of models poses some notoriously difficult numerical challenges. On one hand, Monte Carlo (MC) sampling of the classical field requires repeated diagonalization of the single-particle fermion matrix. On the other hand, it is difficult to eliminate size effects when the effective interactions between the classical degrees of freedom (CDOF) is orders of magnitude smaller than the bare interaction between the fermions and the  CDOF~\cite{WangZ2022a}.
While the first problem has been addressed by different approaches that  reduce the numerical cost of the simulations~\cite{AlonsoJL2001,FurukawaN2004,AlvarezG2007,BarrosK2013}, the second problem, which arises in the weak-coupling to intermediate coupling regime, is more difficult to solve. By ``weak-coupling'' we mean that the value of the interaction $J$ between the classical and fermionic degrees of freedom is small in comparison to the characteristic energy scale of the 
fermions (e.g. the dominant hopping amplitude $t$ of the conduction electrons of a KLM). Since this is the relevant regime for different incarnations of this class of models, such as 4$f$-electron materials described by  KLMs, it is necessary  to develop new approaches that can address this  problem. 

The traditional approach for the weak to intermediate-coupling regimes of this class of models  is perturbation theory. Effective interactions between the CDOF are obtained by expanding them in powers of $J/t$. For instance, the application of perturbation theory to the  KLM leads to the celebrated RKKY interaction between localized moments~\cite{RudermanMA1954,KasuyaT1956,YosidaK1957}. Unfortunately, this
perturbative expansion can not be extended beyond the second order because the corresponding diagrams diverge at $T=0$~\cite{AkagiY2012,OzawaR2016,BatistaCD2016_review,HayamiS2017} , suggesting that the coefficients of  $n$-spin interactions with $n \geq 4$ are non-analytic functions of the coupling constant.  In general, the problem of integrating out fermionic degrees of freedom in presence  of a Fermi surface is highly non-trivial because the effective interactions are expected to be non-analytic functions of the coupling constant. Based on the success of the RKKY model~\cite{RudermanMA1954,KasuyaT1956,YosidaK1957}, here we will conjecture that, for weak-enough coupling constant, one can still neglect $n$-spin interactions with $n >4$ and approximate the 
effective two- and four-spin interactions by analytic functions.
 In other words, we will assume that, despite the non-analytic dependence of the coefficients of $n$-spin interactions on the coupling constant $J$, there is still a hierachy of interactions, i.e., there is a regime where six and higher-spin interactions can be neglected in comparison to the terms including two and four-spin interactions. Moreover, we will also assume that that non-analytic behavior in momentum space, caused by the long-range nature of the real space interactions, can be approximated by a sequence of analytic functions obtained by systematically increasing the range of the interactions. 
As we will discuss in this work, the verification of these conjectures is particularly relevant for addressing situations where the RKKY Hamiltonian is frustrated in the sense that the exchange interaction in momentum space $\tilde{\cal J}(\bm{q})$ is minimized by multiple symmetry related wave-vectors ${\bm Q}_{\nu}$. 

The problem of frustrated magnetic metals has multiple attractive aspects. For instance,  four-spin interactions can stabilize non-coplanar orderings that  induce nonzero Berry curvature of the reconstructed bands. This momentum space Berry curvature can in turn lead to a large topological Hall effect below the magnetic ordering temperature $T_N$~\citep{YeJ1999,OnodaM2004,MartinI2008,YiSD2009,KatoY2010,HamamotoK2015,GobelB2017}. Since $T_N$ can be  comparable or even  higher than room temperature,  multiple  experimental efforts are trying to achieve this goal~\cite{ShaoQ2019}. An outstanding example is the search for  Skyrmion crystals (SkXs) in $f$-electron magnets. Field-induced SkXs with large topological Hall effect have been recently observed in the rare earth based centro-symmetric materials Gd$_2$PdSi$_3$ and Gd$_3$Ru$_4$Al$_{12}$~\cite{MallikR1998a,SahaSR1999,KurumajiT2019,ChandragiriV2016,HirschbergerM2019} that can be modelled by a KLM. 

In a recent work~\cite{WangZ2020}, we have demonstrated that SkXs emerging from triangular lattice RKKY models withe easy-axis anisotropy  can naturally lead to values of the Hall conductivity that are comparable to the quantized value ($e^2/h$).
The key observation is that  the magnitude $Q \equiv |{\bm Q}_{\nu}|$ of magnetic ordering wave vector is  dictated by  the Fermi wave vector $Q \simeq 2k_F$. In a subsequent work~\cite{WangZ2022a}, we used a systematic variational study of the triangular KLM  to show  that  SkXs are  ubiquitous phases of  centrosymmetric metals with localized magnetic moments.  This variational study, which is crucially important to eliminate undesirable finite size effects, reveals mesoscale field-induced  SkXs, whose stability range depends on the coupling strength $J/t$ ($t$ is the nearest-neighbor hopping of the KLM). These results are consistent with an increasing amount of numerical and experimental evidence in favor of the emergence of multi-${\bm Q}$ orderings, including SkXs, in metallic six-fold symmetric layered materials comprising localized magnetic moments coupled via exchange interaction to conduction
electrons~\cite{MallikR1998a,SahaSR1999,KurumajiT2019,ChandragiriV2016,HirschbergerM2019}. 

The emergence of multi-${\bm Q}$ magnetic orderings in Kondo lattice systems has stimulated different groups to propose effective four-spin interactions that can account for this non-linear effect of the conduction electrons. In view of the lack of a controlled analytical procedure, the proposed effective models are mostly phenomenological and ad-hoc because they do not consider all the symmetry allowed four-spin interactions. For instance, Hayami {\it et al.}~\cite{HayamiS2017,HayamiS2021_review} have proposed a phenomenological bilinear biquadratic model based on a trend that they observe in the divergent terms of the perturbative expansion. While phenomenological approaches can offer some useful insights, they severely limit the predictive power of the original high-energy model.  A similar consideration applies to the low-energy excitations (magnons) of each magnetically ordered state. It is then relevant to ask if there is an alternative method that can output the symmetry allowed effective biquadratic spin interactions and simultaneously preserve the predictive power of the original KLM.  A method with these characteristics can be used to understand the origin of the different magnetic orderings induced by magnetic field and/or single-ion anisotropy, as well as to compute the low-energy magnon spectrum of each magnetic phase. As we will see in this work, another important  advantage is that simulations of the effective spin model turn out to be a few orders of magnitude faster  than simulations of the original KLM. 

The alternative approach that we propose here is inspired by a recent proposal for constructing effective low-energy Hamiltonians by supervised learning on energy~\cite{FujitaH2018}. In their work, Fujita et al. used supervised learning to derive an effective spin-1/2 Hamiltonian in the strong-coupling limit of a half-filled Hubbard model. Unlike the case that we consider here, this problem admits a well behaved perturbative expansion, which can be used to corroborate the success of the  supervised learning algorithm. The classical nature of the spin degrees of freedom in the KLM that we consider here introduces another important  difference. The spectrum of classical spin model is always continuous, while the spectrum of a quantum spin model is discrete on finite lattices. In other words, each product of coherent spin states is an eigenstate in the classical limit, while this of course is not true for the quantum case. Consequently, while one can always fit the lowest energy $M$ states of the high-energy model to determine  the optimal parameters of the low-energy quantum spin model, this is not possible for classical spins. As we will see in the next sections, one must introduce an iterative protocol to sample from the continuous set of  low-energy states of the KLM. However, the advantage of the classical case is that the cost function becomes a convex function globally (i.e. it has no local minima), which vastly reduces the cost of the optimization procedure after each iteration.   

From a pure mathematical standpoint, the challenge is to generate a variational low-energy space common to both the original high-energy model ${\mathcal{H}}$ and an effective low-energy spin model $\tilde{\mathcal{H}}$, where the spectrum of ${\mathcal{H}}$ is accurately reproduced. We propose an iterative procedure in which the effective model $\tilde{\mathcal{H}}$  obtained after each iteration is used to update the variational space. 
A key challenge is to find a good balance between the relative weights assigned to low and high-energy states of the original high-energy model ${\mathcal{H}}$. Another challenge is to find an algorithm that converges relatively fast to the final version of $\tilde{\mathcal{H}}$.
As we will see in the following sections, the algorithm that we are proposing meets both challenges. The main limiting factor for the efficiency of the algorithm  is the time associated with the generation of the initial training data set. Sections~\ref{HEM} and ~\ref{LEM} introduce the high and low-energy models that we use to test the proposed algorithm. The high-energy model is a triangular KLM, whose filling fraction is tuned to favor periodic magnetic structures with a magnetic unit cell of $6 \times 6$ spins. The low-energy model contains the usual bilinear RKKY interactions~\cite{RudermanMA1954,KasuyaT1956,YosidaK1957}, plus all the symmetry allowed four-spin interactions up to a certain range in real space. The proposed algorithm is introduced in Sec.~\ref{MLA}. The generation of the training data set is described in Sec.~\ref{KLM}. Section~\ref{results} includes a comparison between the zero temperature phase diagrams of the original KLM and the one obtained with the low-energy model. The effective low-energy spin model $\tilde{\mathcal{H}}$ is further validated in Sec.~\ref{DYN}, which includes a comparison of the dynamical spin structure factor in the fully polarized phase computed with ${\mathcal{H}}$ and $\tilde{\mathcal{H}}$. The comparison reveals a good agreement for the magnon dispersion, which was not used as part of the training data set. The final Sec.~\ref{S&O} includes a summary and outlook of the results presented in this work.



\section{High Energy Model \label{HEM}}

The discovery of magnetic SkX 
in chiral magnets, such as MnSi, Fe$_{1-x}$Co$_x$Si, FeGe and Cu$_2$OSeO$_3$~\citep{MuhlbauerS2009, YuXZ2010,YuXZ2011,SekiS2012,AdamsT2012} 
spawned efforts for identifying stabilization mechanisms of SkX in different classes of materials. These efforts are revealing that  new stabilization mechanisms are typically accompanied by novel physical  properties. For instance, while the vector chirality is fixed in the magnetic skyrmions of chiral magnets such as the above-mentioned B20 compounds, it is a degree of freedom in the SkX of centrosymmetric materials such as BaFe$_{1-x-0.05}$Sc$_x$Mg$_{0.05}$O$_{19}$, La$_{2-2x}$Sr$_{1+2x}$Mn$_2$O$_7$, Gd$_2$PdSi$_3$ and Gd$_3$Ru$_4$Al$_{12}$~\cite{YuX2012,YuXZ2014,MallikR1998a,SahaSR1999,KurumajiT2019,ChandragiriV2016,HirschbergerM2019}. In the former case, the underlying spiral structure emerges from the  competition between ferromagnetic exchange and the Dzyaloshinskii-Moriya (DM) interaction~\cite{DzyaloshinskyI1958,MoriyaT1960a}. In contrast, the spiral ordering of centrosymmetric materials arises from frustration, i.e., from the competition between different exchange  or dipolar interactions~\citep{OkuboT2012,LeonovAO2015,LinSZ2016a,HayamiS2016,BatistaCD2016_review}.

Most of the known magnetic SkXs have been reported for metallic materials, where the interplay between magnetic moments and conduction electrons leads to 
response functions that are of both fundamental and applied interests, such as the well-known topological Hall effect (THE)~\citep{OnodaM2004,YiSD2009,HamamotoK2015,GobelB2017} and the current-induced skyrmion motion~\citep{JonietzF2010,YuXZ2012,SchulzT2012,NagaosaN2013}. The THE is a direct consequence of the Berry curvature acquired by the reconstructed electronic bands. In the adiabatic limit, the momentum space Berry curvature is controlled by a real space Berry curvature, that is proportional to the skyrmion density in absence of spin-orbit interaction~\cite{Zhang20}: each skyrmion produces an effective flux equal to the flux quantum $\Phi_0$. Consequently, Hall conductivities comparable to the quantized value ($e^2/h$)  can in principle be achieved if the ordering wave vector of the SkX is comparable to the Fermi wave vector $k_F$. This condition can be naturally fulfilled in $f$-electron systems where the  interaction between magnetic moments is mediated by conduction electrons~\cite{WangZ2020}. 
Indeed, field-induced SkXs with large topological Hall effect have been recently observed in the rare earth based centrosymmetric materials Gd$_2$PdSi$_3$ and Gd$_3$Ru$_4$Al$_{12}$~\cite{MallikR1998a,SahaSR1999,KurumajiT2019,ChandragiriV2016,HirschbergerM2019}, which are in principle described by a KLM including multiple bands of conduction electrons coupled to  localized $f$-magnetic moments. The combination of state of the art band structure calculations with standard degenerate perturbation theory seems to be a promising route for deriving these KLMs from first principle calculations~\cite{Simeth22}. Consequently, it is important to develop efficient methodologies for computing the quantum ($T=0$) phase diagram of this class of models.


To understand the origin of the field-induced SkX phases, we will consider a simple triangular KLM with  \emph{classical} local magnetic moments:
\begin{align}
\mathcal{H}&= -t \sum_{\langle \bm{r},\bm{r}^\prime \rangle,\sigma}  \left(c_{\bm{r}\sigma}^{\dagger}c_{\bm{r}^\prime\sigma}+h.c.\right) +J\sum_{\bm{r} ,\alpha\beta }c_{ \bm{r}\alpha}^{\dagger}\bm{\sigma}_{\alpha\beta}c_{\bm{r} \beta}\cdot\bm{S}_{\bm{r}} 
\nonumber \\
&\quad - h\sum_{\bm{r}}S_{\bm{r}}^{z}+D\sum_{\bm{r}}\left(S_{\bm{r}}^{z}\right)^{2},
\label{eq:Kondo}
\end{align}
where the first term corresponds to a tight-binding model with  the nearest-neighbor hopping amplitude $t$ ($\langle \bm{r},\bm{r}^\prime \rangle$ denotes bonds of nearest-neighbor sites). The operator $c_{\bm{r}\sigma}^{\dagger}$ ($c_{\bm{r}\sigma}$) creates (annihilates)
an itinerant electron with spin $\sigma$ on site $\bm{r}$. 
$J$ is the Kondo exchange interaction between the local magnetic moments $\bm{S}_{\bm{r}}$
and the conduction electrons ($\bm{\sigma}$ is the vector of the
Pauli matrices), 
and the localized moments are described by the normalized classical vector field $\bm{S}_{\bm{r}}$ ($\left|\bm{S}_{\bm{r}}\right|=S$).~\footnote{In this work we simply set $S=1$.} The last two terms represent a Zeeman coupling to an external field $H$ ($h = g \mu_B H$) and an easy-axis single-ion anisotropy ($D<0$). 

The coupling of localized spins with  itinerant electrons  leads to effective spin-spin interactions which can potentially stabilize multiple-$\bm{Q}$ magnetic orderings.
Modeling the system with the ``high-energy'' KLM is numerically challenging because of a combination of reasons.
The numerical cost of computing energy for a lattice of $N$
sites is of the order of $\mathcal{O}(N^{3})$. While this cost can in principle be reduced to a linear function of $N$ by implementing approximated numerical schemes, 
such as the Kernel Polynomial Method (KPM)~\cite{WeisseA2006_RMP,BarrosK2013,WangZ2018}, the problem that still persists is that for most instances finite-size effects remain relevant even for very large lattices~\cite{WangZ2022a}. This important limitation was overcome by a variational calculation for the case of periodic structures with relatively small periods, which are fixed by carefully tuning the band filling fraction.
While these approaches can be used to obtain zero-temperature phases diagrams, like the ones that we will discuss in Sec.~\ref{results}, they cannot be used to study dynamical response function or the finite-temperature phase diagram. Moreover, it is difficult to extract stabilization mechanisms of multi-{${\bm Q}$} orderings from numerical solutions of the KLM. Since these orderings can be determined by $n$-spin interactions with $n\geq 4$, the derivation of a low-energy model beyond the RKKY level is crucial to understand  different aspects of the  competition between single and multi-{${\bm Q}$} orderings.

\begin{figure*}
\includegraphics[scale=0.4]{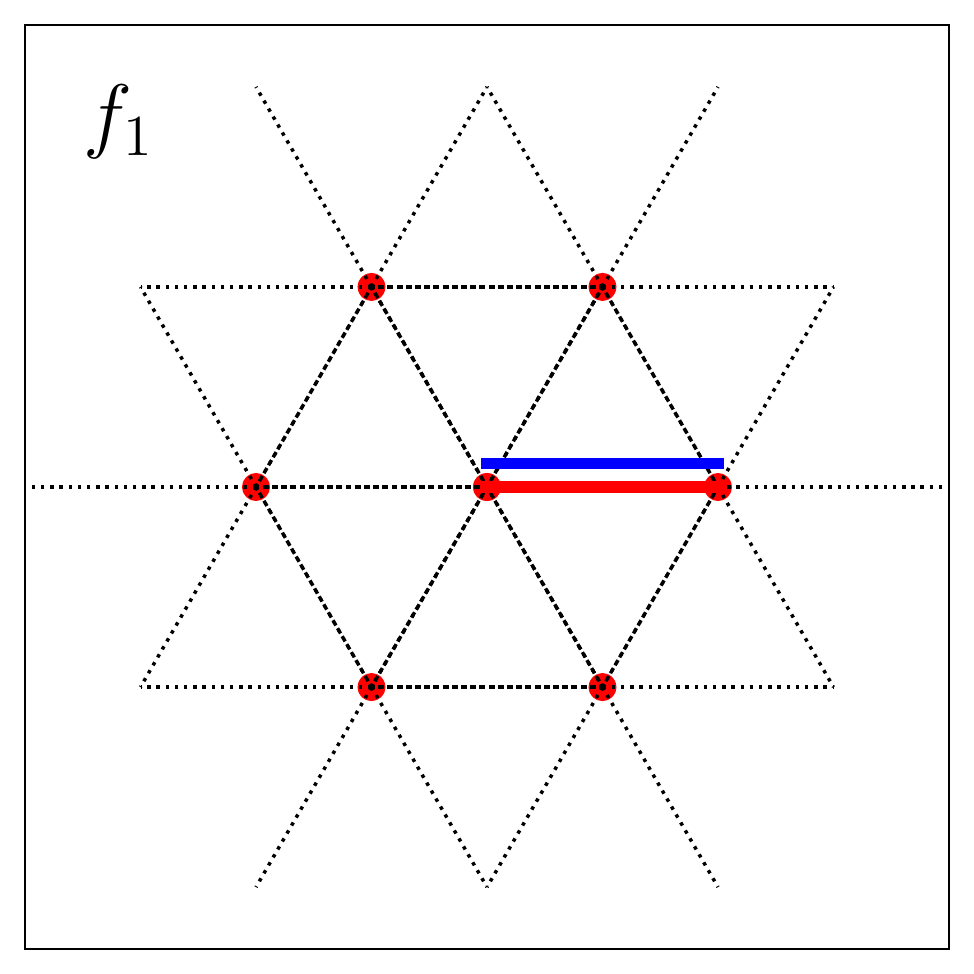}
\includegraphics[scale=0.4]{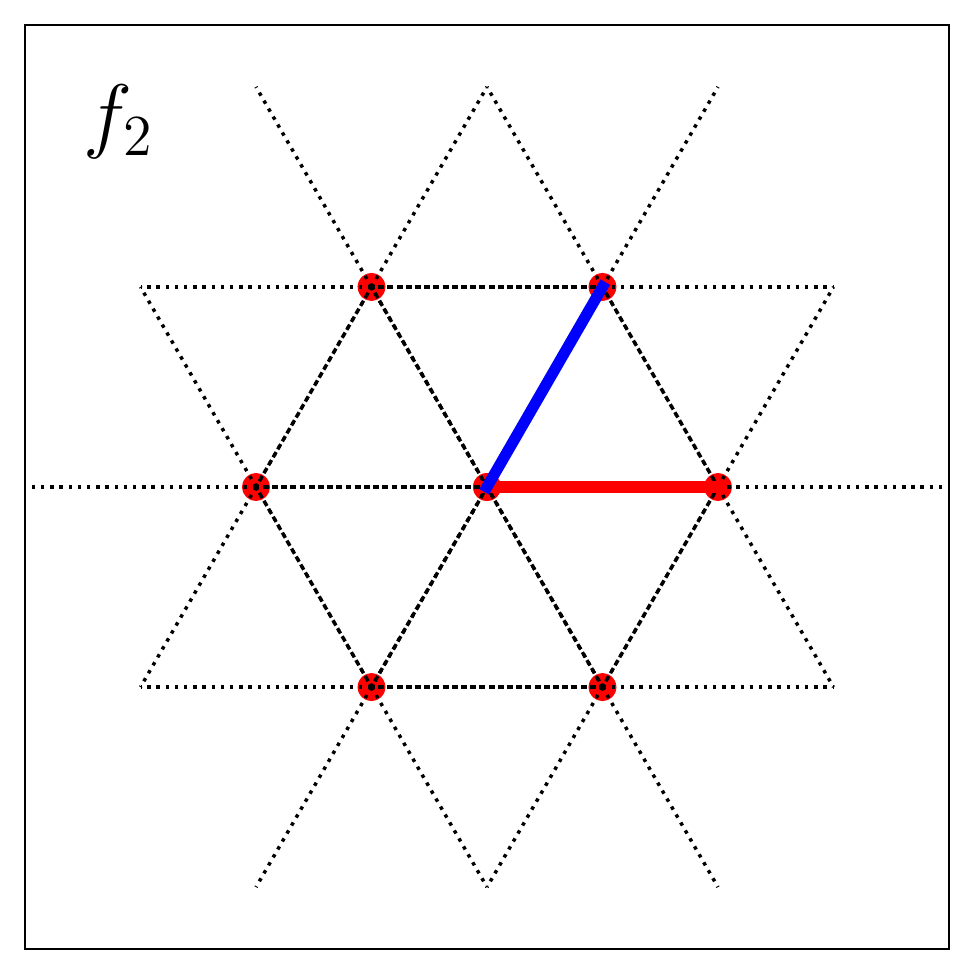} \includegraphics[scale=0.4]{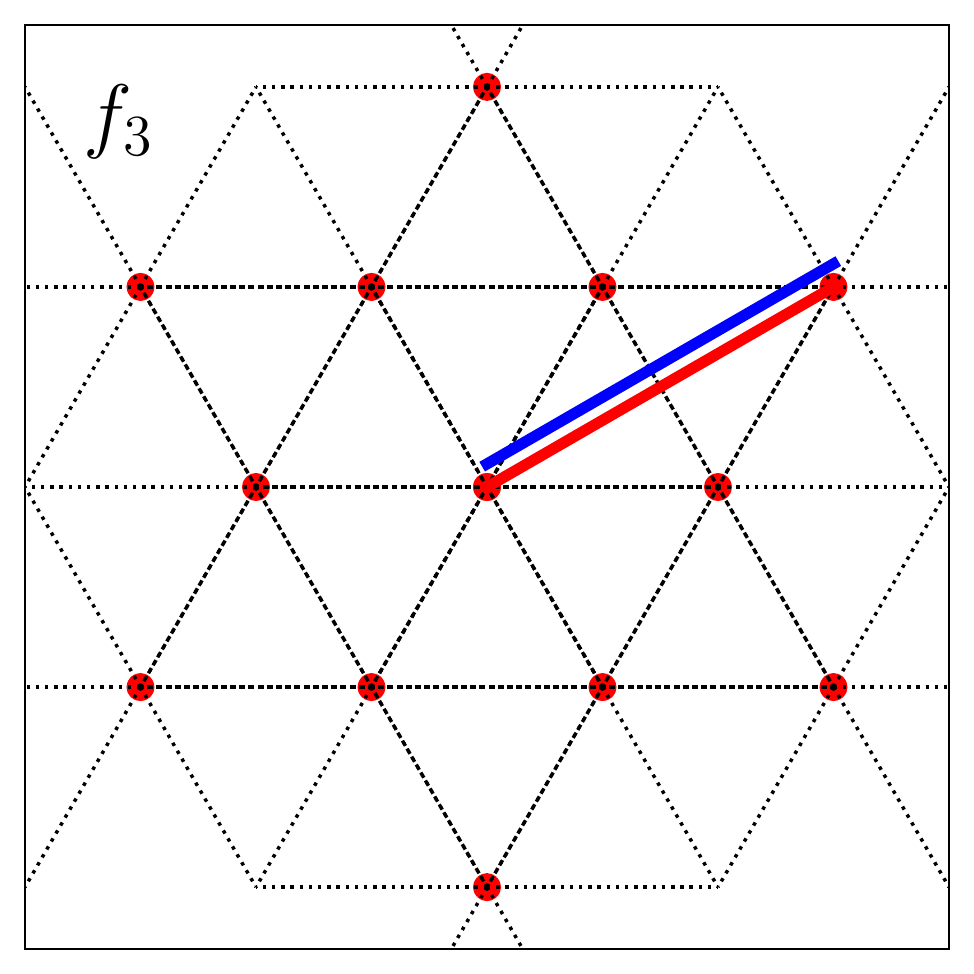}
\includegraphics[scale=0.4]{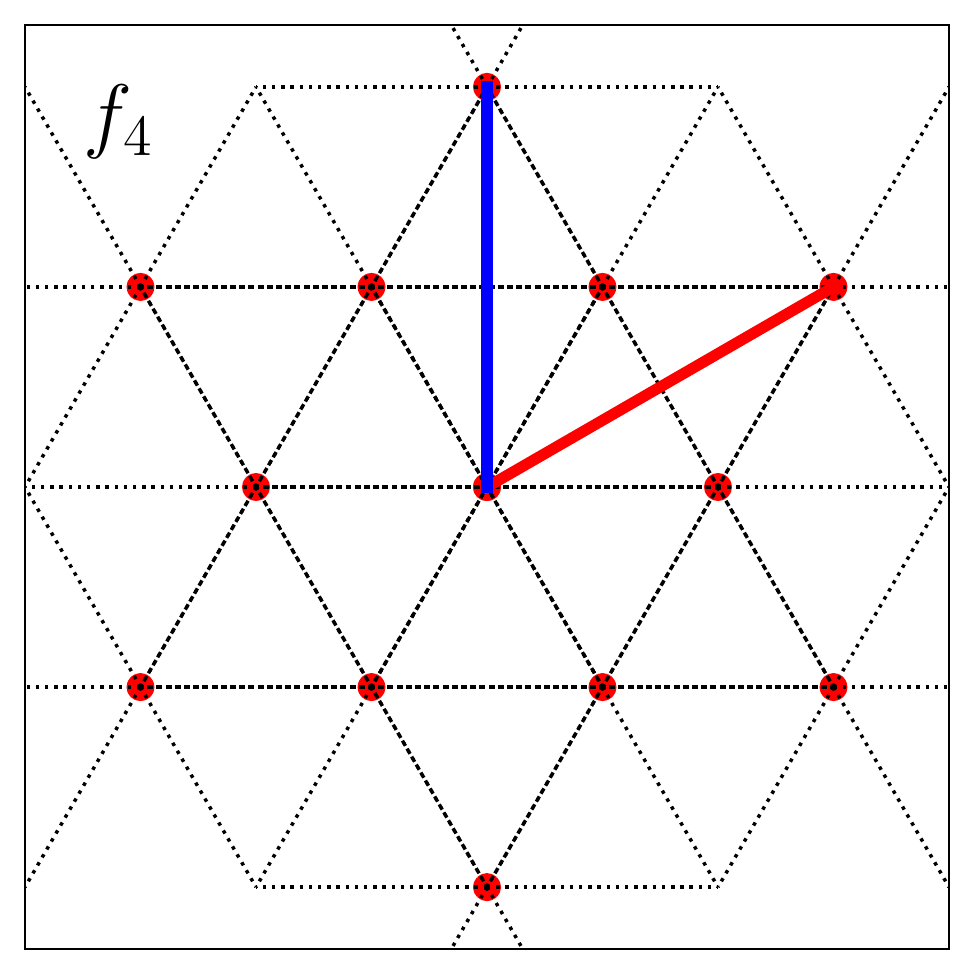}

\includegraphics[scale=0.4]{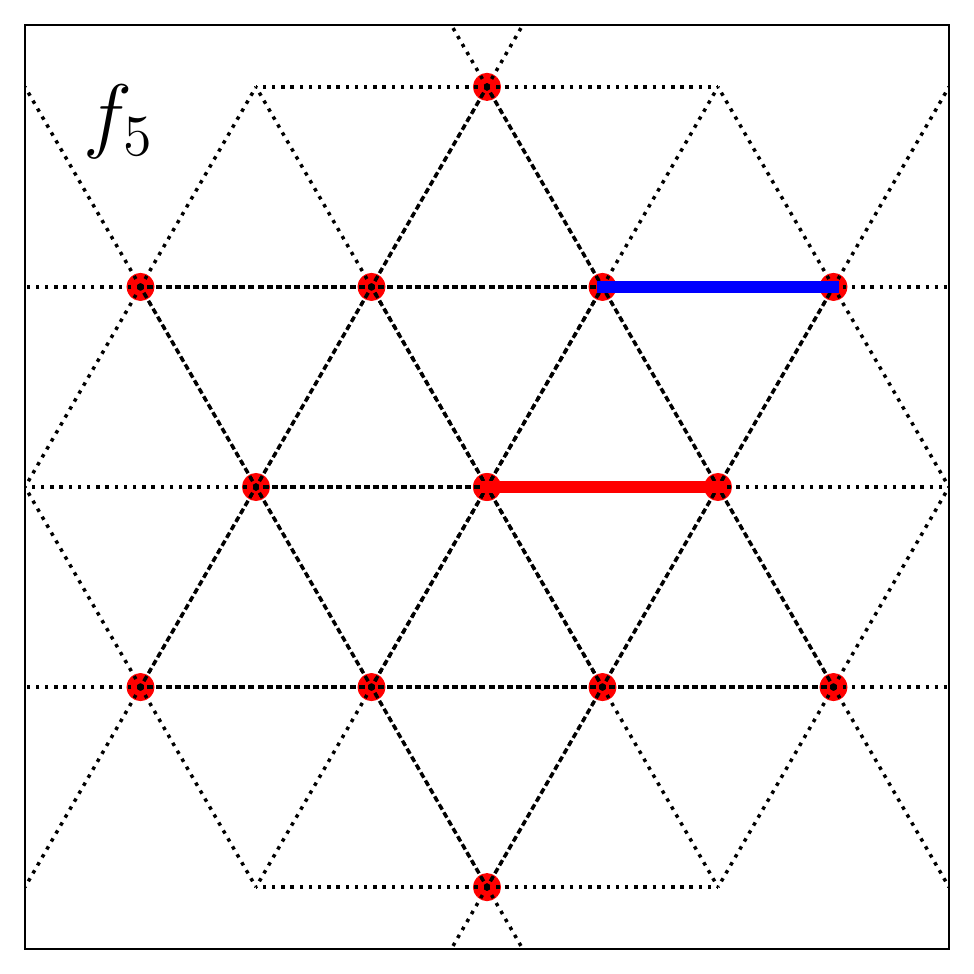}
\includegraphics[scale=0.4]{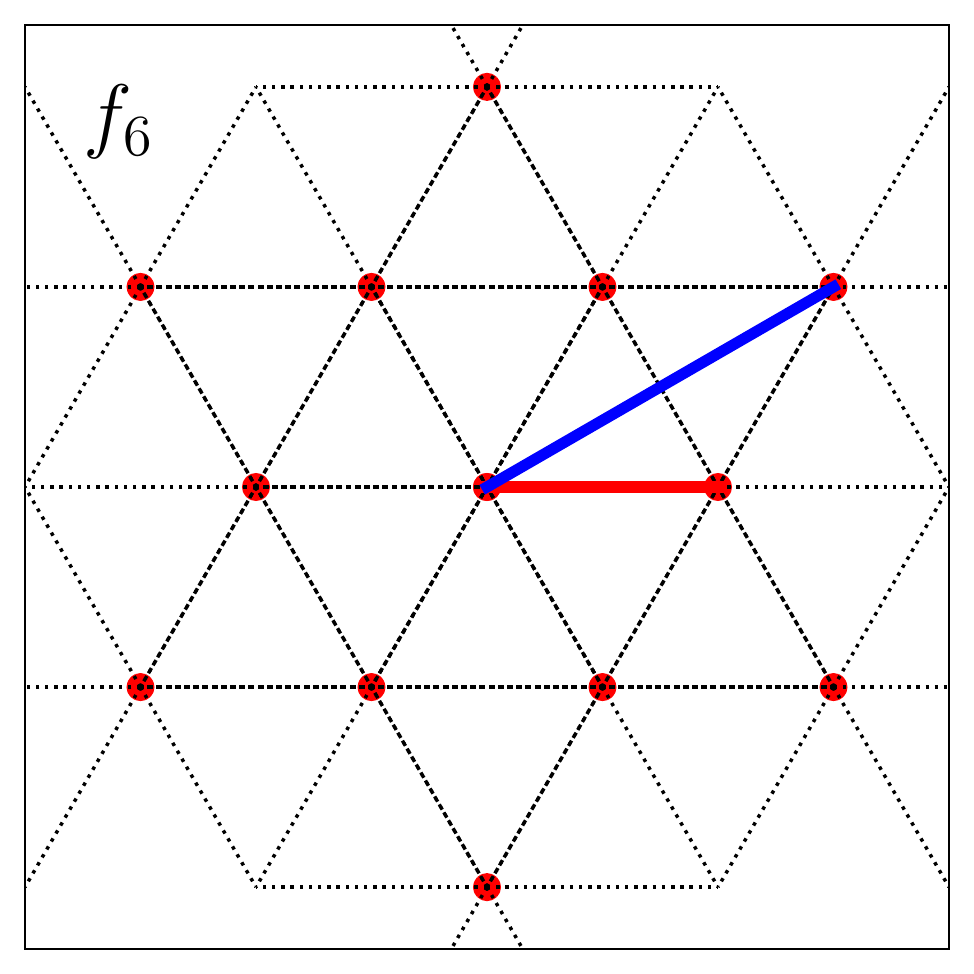} \includegraphics[scale=0.4]{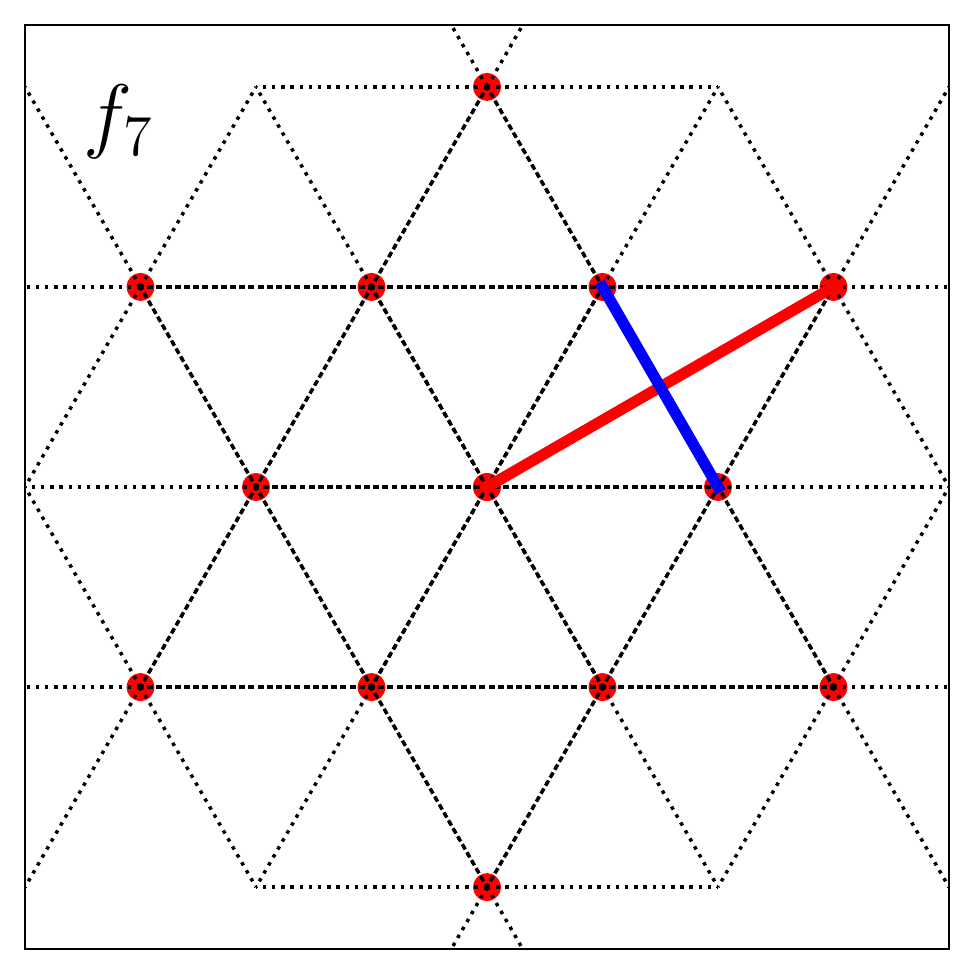}
\caption{Four-spin interactions with cutoff $|\bm{r}_i - \bm{r}_j| \le \sqrt{3}a$ on a triangular lattice. The red and blue lines represent  $\left(\bm{S}_{\bm{r}_1}\cdot \bm{S}_{\bm{r}_2}\right)$ and $\left(\bm{S}_{\bm{r}_3}\cdot \bm{S}_{\bm{r}_4}\right)$ terms respectively. }
\label{fig:four_spin_interactions}
\end{figure*}

\section{Low energy model \label{LEM}}
In the weak-coupling limit, $J\eta(\epsilon_{F})\ll1$ , where $\eta(\epsilon_{F})$
is the density of states at Fermi energy $\epsilon_{F}$,  the KLM can be approximated by
an effective  RKKY model~\cite{RudermanMA1954,KasuyaT1956,YosidaK1957}:
\begin{equation}
\mathcal{H}_{\rm RKKY}=\sum_{\bm{k}}\tilde{\cal J}(\bm{k})\bm{S}_{\bm{k}} \cdot \bm{S}_{\bm{-k}}\label{eq:RKKY},
\end{equation}
where 
\begin{equation}
\bm{S}_{\bm{k}}=\frac{1}{\sqrt{N}}\sum_{\bm{r}}e^{-\iu \bm{k} \cdot \bm{r}}\bm{S}_{\bm{r}}
\end{equation}
is the Fourier transform of the spins, and $N$ is the total number of lattice sites.
In momentum space,  the effective RKKY interaction is proportional to the magnetic susceptibility $\chi_{\bm{k}}$ of the conduction
electrons:
\begin{equation}
\tilde{\cal J}(\bm{k})=-J^{2}\chi_{\bm{k}},
\end{equation}
where 
\begin{equation}
\chi_{\bm{k}}=-\int \frac{d\bm{q}}{\mathcal{A}_\mathcal{B}} \frac{f\left(\epsilon_{\bm{q}+\bm{k}}\right)-f\left(\epsilon_{\bm{q}}\right)}{\epsilon_{\bm{q}+\bm{k}}-\epsilon_{\bm{q}}}.
\label{eq:magnetic_sus}
\end{equation}
Here $f\left(\epsilon\right)$ is the Fermi distribution function, $\mathcal{A}_\mathcal{B}$ is the area of the first Brillouin zone, and $\epsilon_{\bm{k}}$ is the bare dispersion of the electrons. Since $\bm{S}_{\bm{k}} = \bm{S}^*_{-\bm{k}}$ ($\bm{S}_{\bm{r}}$ is a real vector field) implies that $\bm{S}_{\bm{k}}\cdot \bm{S}_{\bm{-k}} \geq 0$
and
$
\sum_{\bm k}  \bm{S}_{\bm{k}}\cdot \bm{S}_{\bm{-k}} = N S^2,
$
the RKKY interaction favors a helical spin ordering with an ordering wave factor that minimizes $\tilde{\cal J}(\bm{k})$ (maximizing $\chi_{\bm{k}}$). If a wave vector $\bm{Q}$ that optimizes the RKKY model~\eqref{eq:RKKY} is not invariant under the symmetry group ${\cal G}$ of transformations that leave the KLM invariant, there are other wave vectors related by symmetry operations ${\cal G}\bm{Q}$ that also minimize the energy. 
The  symmetry related wave vectors will be denoted with the index $\nu$: $\{ \bm{Q}_{\nu} \}$. The spatial inversion operation is not included in the symmetry group because, as we mentioned above, the $\bm{Q}_{\nu}$ and $-\bm{Q}_{\nu}$ components of the vector field are not independent ($\bm{S}_{\bm{Q}_{\nu}} = \bm{S}^*_{-\bm{Q}_{\nu}}$). 
Thus, there is a degeneracy for the helical orderings of symmetry related wave vectors. The real space version of the RKKY model is given by Fourier transforming Eq.~\eqref{eq:RKKY}:
\begin{equation}
\mathcal{H}_{\rm RKKY}=\frac{1}{2} \sum_{ \bm{r} \neq \bm{r}' } {\cal J}(\bm{r}-\bm{r}')\bm{S}_{\bm{r}}\cdot \bm{S}_{\bm{r}'}.\label{eq:RKKY_real}
\end{equation}

Away from the weak coupling limit, four-spin and higher order interactions
are naturally generated from the KLM, which are ignored in the RKKY model.
The four-spin interactions are vital in getting a field-induced SkX
phase at zero anisotropy, since the RKKY part alone is not sufficient
to stabilize this ordering. 
\begin{align}
\mathcal{H}_{\rm eff}  &=  \sum_{\bm{k}}\tilde{\cal J}(\bm{k})\bm{S}_{\bm{k}}\cdot \bm{S}_{\bar {\bm{k}}} \nonumber \\
 &\quad +  \sum_{\bm{k}_{1},\bm{k}_{2},\bm{k}_{3}}  \frac{g\left(\bm{k}_{1},\bm{k}_{2},\bm{k}_{3}\right)}{N}\left(\bm{S}_{-{\bm K}}\cdot \bm{S}_{\bm{k}_{1}}\right)  \left(\bm{S}_{\bm{k}_{2}}\cdot \bm{S}_{\bm{k}_{3}}\right),
\label{eq:model_ham_mom}
\end{align}
where 
${\bm K} \equiv \bm{k}_{1} + \bm{k}_{2}+ \bm{k}_{3}$.
The four-spin interaction term helps to lift the massive ground state degeneracy of the RKKY model.
Suppose the term $\left(\bm{S}_{\bm{Q}_{i}}\cdot \bm{S}_{\bm{-Q}_{i}}\right)\left(\bm{S}_{\bm{Q}_{i}}\cdot \bm{S}_{\bm{-Q}_{i}}\right)$
in the Hamiltonian has a large positive coefficient $g\left(-\bm{Q}_{i},\bm{Q}_{i},\bm{-Q}_{i}\right)$,
then the single-$\bm{Q}_{i}$ ordering will be heavily penalized. On the other hand, a large negative coefficient $g(\bm{Q}_1,\bm{Q}_2,\bm{Q}_3)$ favors three different $\bm{Q}$
components $\left(\bm{Q}_{1},\bm{Q}_{2},\bm{Q}_{3}\right)$
being finite in the presence of a finite uniform magnetization $\bm{S}_{0}$.
In other words, this term favors a triple-$\bm{Q}$ ordering,
such as SkXs, if a finite magnetization is induced by
an external magnetic field. 
Thus, the function $g(\bm{k}_1,\bm{k}_2,\bm{k}_3)$ plays a critical role in the identification of the ground state. As we demonstrate in this paper, ML turns out to be a valuable tool in extracting this important piece of information.

{Since $g\left(\bm{k}_{1},\bm{k}_{2},\bm{k}_{3}\right)$
is an unknown function of 3 continuous multidimensional variables, we need an efficient scheme to reduce the number of model
parameters.
To obtain a valid effective low-energy theory, the scheme should  provide a good approximation 
near the most relevant wave vectors ($\bm{k}\simeq\bm{Q}_{\nu}$
and $\bm{k}\simeq\bm{0}$). 
To achieve this goal, we will first derive a real space version
of the low-energy effective Hamiltonian, where all symmetry allowed interactions are included    up to a certain distance. 
After these real space interactions
are obtained with ML, the expressions of $\tilde{\cal J}(\bm{k})$ and $g\left(\bm{k}_{1},\bm{k}_{2},\bm{k}_{3}\right)$ can be obtained by a simple Fourier transform of the effective real space interactions. }

For the RKKY contribution~\eqref{eq:RKKY_real}, we cut off the interactions beyond $|\bm{r} - \bm{r}'| = 2\sqrt{3}a$, which leads to 6 inequivalent exchange parameters \{$\mathcal{J}_{1}$, \ldots,  $\mathcal{J}_{6}$\}.
Similarly,  the four-spin contribution can be expressed as:
\begin{equation}
\mathcal{H}_{4\text{-spin}}=\sum_{\langle \bm{r}_1,\bm{r}_2,\bm{r}_3,\bm{r}_4 \rangle}f(\bm{r}_1,\bm{r}_2,\bm{r}_3,\bm{r}_4) \left(\bm{S}_{\bm{r}_1}\cdot \bm{S}_{\bm{r}_2}\right)\left(\bm{S}_{\bm{r}_3}\cdot \bm{S}_{\bm{r}_4}\right),
\label{eq:4spin_real}
\end{equation}
where the notation $\langle \bm{r}_1,\bm{r}_2,\bm{r}_3,\bm{r}_4 \rangle$ indicates that each set of 
four sites ($\bm{r}_1,\bm{r}_2,\bm{r}_3,\bm{r}_4$) is counted only once.
Here we restrict 
$\vert\bm{r}_{i}-\bm{r}_{j}\vert\le \sqrt{3}a$,  which leads to 7 inequivalent exchange parameters \{$f_1$, \ldots, $f_7$\} (see Fig.~\ref{fig:four_spin_interactions}).

By implementing the ML algorithm described in the next section, we compute these real space parameters and subsequently evaluate $\tilde{\cal J}(\bm{k})$ and $g\left(\bm{k}_{1},\bm{k}_{2},\bm{k}_{3}\right)$.  By limiting the possible $\bm{k}$-values to the set  $\left(\bm{Q}_{1},\bm{Q}_{2},\bm{Q}_{3},\bm{0}\right)$,
the energy contribution from the four-spin interactions can be written
as:
\begin{widetext}
\begin{align}
NE_{4\text{-spin}} & =\tilde{g}_{0}\left(\bm{S}_{\bm{0}}\cdot \bm{S}_{\bm{0}}\right)^2 + \tilde{g}_{1}\sum_{\nu} \left(\bm{S}_{\bm{0}} \cdot \bm{S}_{\bm{0}}\right) \left(\bm{S}_{\bm{Q}_{\nu}} \cdot \bm{S}_{-\bm{Q}_{\nu}}\right) +\tilde{g}_{2}\sum_{\nu} \left(\bm{S}_{\bm{0}}\cdot \bm{S}_{\bm{Q}_{\nu}}\right) \left(\bm{S}_{\bm{0}} \cdot \bm{S}_{-\bm{Q_{\nu}}} \right) \nonumber \\
&\quad+\tilde{g}_{3}  \left[ \left(\bm{S}_{\bm{0}}\cdot \bm{S}_{\bm{Q}_{1}}\right)\left(\bm{S}_{\bm{Q}_{2}}\cdot \bm{S}_{\bm{Q}_{3}}\right)+\left(\bm{S}_{\bm{0}}\cdot \bm{S}_{-\bm{Q}_{1}}\right)\left(\bm{S}_{-\bm{Q}_{2}} \cdot \bm{S}_{-\bm{Q}_{3}}\right) \right] 
\nonumber \\
&\quad+\tilde{g}_{3}  \left[ \left(\bm{S}_{\bm{0}}\cdot \bm{S}_{\bm{Q}_{2}}\right)\left(\bm{S}_{\bm{Q}_{1}}\cdot \bm{S}_{\bm{Q}_{3}}\right)+\left(\bm{S}_{\bm{0}}\cdot \bm{S}_{-\bm{Q}_{2}}\right)\left(\bm{S}_{-\bm{Q}_{1}} \cdot \bm{S}_{-\bm{Q}_{3}}\right) \right] 
\nonumber \\
&\quad+\tilde{g}_{3}  \left[ \left(\bm{S}_{\bm{0}}\cdot \bm{S}_{\bm{Q}_{3}}\right)\left(\bm{S}_{\bm{Q}_{1}}\cdot \bm{S}_{\bm{Q}_{2}}\right)+\left(\bm{S}_{\bm{0}}\cdot \bm{S}_{-\bm{Q}_{3}}\right)\left(\bm{S}_{-\bm{Q}_{1}} \cdot \bm{S}_{-\bm{Q}_{2}}\right) \right] 
\nonumber \\
& \quad+\tilde{g}_{4}\left[ \left(\bm{S}_{\bm{Q}_{1}}\cdot \bm{S}_{-\bm{Q}_{2}}\right) \left(\bm{S}_{-\bm{Q}_{1}} \cdot \bm{S}_{\bm{Q}_{2}}\right)+\left(\bm{S}_{\bm{Q}_{2}}\cdot \bm{S}_{-\bm{Q}_{3}}\right) \left(\bm{S}_{-\bm{Q}_{2}}\cdot \bm{S}_{\bm{Q}_{3}}\right)
+\left(\bm{S}_{\bm{Q}_{3}}\cdot \bm{S}_{-\bm{Q}_{1}}\right)\left(\bm{S}_{-\bm{Q}_{3}}\cdot \bm{S}_{\bm{Q}_{1}}\right)\right]\nonumber \\
& \quad+\tilde{g}_{5}\left[ \left(\bm{S}_{\bm{Q}_{1}}\cdot \bm{S}_{\bm{Q}_{2}}\right) \left(\bm{S}_{-\bm{Q}_{1}}\cdot \bm{S}_{-\bm{Q}_{2}}\right) +\left(\bm{S}_{\bm{Q}_{2}}\cdot \bm{S}_{\bm{Q}_{3}}\right) \left(\bm{S}_{-\bm{Q}_{2}}\cdot \bm{S}_{-\bm{Q}_{3}}\right) + \left(\bm{S}_{\bm{Q}_{3}}\cdot \bm{S}_{\bm{Q}_{1}}\right)\left(\bm{S}_{-\bm{Q}_{3}}\cdot \bm{S}_{-\bm{Q}_{1}}\right) \right]\nonumber \\
& \quad+\tilde{g}_{6} \sum_{\nu }\left(\bm{S}_{\bm{Q}_{\nu}}\cdot \bm{S}_{\bm{Q}_{\nu }}\right) \left(\bm{S}_{-\bm{Q_{\nu}}}\cdot \bm{S}_{-\bm{Q_{\nu}}}\right) +\tilde{g}_{7}\sum_{\nu} \left(\bm{S}_{\bm{Q}_{\nu}}\cdot \bm{S}_{-\bm{Q}_{\nu}}\right) \left(\bm{S}_{\bm{Q_{\nu}}}\cdot \bm{S}_{-\bm{Q_{\nu}}}\right) \nonumber \\
& \quad+\tilde{g}_{8}\left[ \left(\bm{S}_{\bm{Q}_{1}}\cdot \bm{S}_{-\bm{Q}_{1}}\right)\left(\bm{S}_{\bm{Q}_{2}}\cdot \bm{S}_{-\bm{Q}_{2}}\right) + \left(\bm{S}_{\bm{Q}_{2}}\cdot \bm{S}_{-\bm{Q}_{2}}\right) \left(\bm{S}_{\bm{Q}_{3}}\cdot \bm{S}_{-\bm{Q}_{3}}\right) + \left(\bm{S}_{\bm{Q}_{3}}\cdot \bm{S}_{-\bm{Q}_{3}}\right)\left(\bm{S}_{\bm{Q}_{1}}\cdot \bm{S}_{-\bm{Q}_{1}}\right) \right] ,
\label{eq:4spin_g_energy}
\end{align}
\end{widetext}
where 
the expressions of $\{\tilde{g}_{i}\}$ are given in Appendix~\ref{sec:g_expression}.

\section{Machine learning algorithm \label{MLA}}

The first step to extract the spin Hamiltonian from the KLM (or any other high energy model) is generating an initial high energy dataset on a grid of magnetic field strength $\left(h\right)$ and easy axis anisotropy $\left(D\right)$. This grid need not be very dense. In particular, our grid for the ML algorithm had 47 points, while the grid for producing the final phase diagram using the low-energy model used more than 3000 points. For each value of $h$ and $D$ on the grid, we generate a random spin configuration. Starting from this random configuration, we find the local minimum using a gradient-based method of the original KLM (calculation details are given in next section). All the spin configurations generated in the minimization process are stored along with their corresponding energies and values of $h$ and $D$ in a dataset~\cite{WangZ2022a}. The next step is to set up a trial Hamiltonian:
\begin{equation}
\mathcal{H}=\sum_{j=1}^{M}c_{j}\mathcal{H}_{j}\qquad(j=1, \ldots, M).
\end{equation}
where ${\mathcal{H}_j}$ consists of a constant term, real space RKKY exchange interactions, and four-spin interactions ( described in previous section). 
The effective low-energy model is expected to reproduce the low-energy spectrum of the original KLM. Correspondingly, to formulate the search of this effective model as an optimization problem, we introduce the cost function:
\begin{equation}
\text{Cost}\left(\left\{ c_{j}\right\} \right)=\frac{1}{\tilde{N}}\sum_{D,h}\sum_{i=0}^{\tilde{N}-1}\delta_{i}^{D,h}\left(E_{i}^{D,h}-\tilde{E}_{i}^{D,h}\right)^{2},\label{eq:CF}
\end{equation}
\begin{equation}
\delta_{i}^{D,h}=\frac{1}{\sqrt{E_{i}^{D,h}-E_\text{min}^{D,h}-\epsilon}},\label{eq:delta}
\end{equation}
where $\tilde{N}$ is the total number of states in the dataset, $E_{i}^{D,h}$ is the energy obtained by the high energy model
for the $i$-th spin configuration for a particular point on grid
$\left(D,h\right)$, $\tilde{E}_{i}^{D,h}$ is the predicted energy
from the trial Hamiltonian, $\epsilon$ is a very small hyperparameter ensuring
that the weight function $\delta_{i}^{D,h}$ does not diverge and $E_\text{min}^{D,h}$
is the lowest energy in the training dataset for a particular value
of $\left(D,h\right)$.

The dataset comprises of considerable number of states stretching
across the full energy spectrum. 
We note that including high-energy states is important 
because in absence of those, 
the training can lead to a model 
which includes some high-energy states in its low-energy spectrum.
The factor $\delta_{i}^{D,h}$ determines the relative weight of low and high-energy states of the original model.
To optimize the weight factor,
we implemented our algorithm with three different choices: 
\begin{equation}
\delta_{i}^{D,h} = \left\{ \frac{1}{\sqrt{E_{i}^{D,h}-E_{min}^{D,h}-\epsilon}}, 
\, 1, 
\,
\frac{1}{\vert E_{i}^{D,h}-E_{min}^{D,h}-\epsilon\vert} \right \}.
\end{equation}
For $\delta_{i}^{D,h}=1$, the precision in the energy prediction for low
energy states was comparable to the energy scale of the low energy
excitations and hence, it was not adequate to establish the appropriate
energy order of the competing low energy states. For the choice 
$$\delta_{i}^{D,h}=\frac{1}{\vert E_{i}^{D,h}-E_{min}^{D,h}-\epsilon\vert},$$ 
there was an excessive weight on the low energy states and, as a
consequence, the trained model had exceedingly depleted precision
for other states. 
The choice 
$$\delta_{i}^{D,h}=\frac{1}{\sqrt{E_{i}^{D,h}-E_{min}^{D,h}-\epsilon}}$$
provided an appropriate balance between the accuracy for the low-energy and high-energy states.

After setting up the cost function and initializing the coupling constants,  we perform the following three steps iteratively:

\begin{enumerate}

\item Update parameters $\left\{c_{j}\right \}$ iteratively using the gradient
descent method: $c_{j}\rightarrow c_{j}-\alpha\frac{\partial\text{Cost }\left(\left\{ c_{j}\right\} \right)}{\partial c_{j}}$
until cost function reaches a minimum ($\alpha$ is the learning rate).
Keeping the learning rate $\alpha$ small, the gradient can be analytically
calculated.
\begin{equation}
c_{j}\rightarrow c_{j}-\alpha\sum_{i=0}^{\tilde{N}-1}\frac{\partial\text{Cost }\left(\left\{ c_{j}\right\} \right)}{\partial E_{i}\left(\left\{ c_{j}\right\} \right)}\frac{\partial E_{i}\left(\left\{ c_{j}\right\} \right)}{\partial c_{j}}
\end{equation}
\begin{equation}
\implies c_{j}\rightarrow c_{j}-\alpha\sum_{i=0}^{\tilde{N}-1}\frac{\partial\text{Cost }\left(\left\{ c_{j}\right\} \right)}{\partial E_{i}\left(\left\{ c_{j}\right\} \right)}\langle\psi_{i}\vert H_{j}\vert\psi_{i}\rangle
\end{equation}
One of the advantages of the search for a low-energy \emph{classical model} is that  multivariate weighted linear regression can be used to find the optimal model
parameters. In contrast, a non-linear regression is required for quantum mechanical low-energy models because the eigenstates change with the model parameters~\cite{FujitaH2018}.
\item  Produce the zero temperature phase
diagram of the ``ML model'' (with the parameters that minimize the cost function) by energy minimization
via a gradient descent method.
For local minimization algorithms, the
converged results are usually metastable local minima, i.e., different initial
spin configurations can lead to different final states. Correspondingly, for each $h$
and $D$, we typically performed 60 independent runs with different
random initial spin configurations to find the global minima.
\item For the global minimum energy states of the ML model 
calculate the KLM energies and add them to the training dataset. Then  update the
weights $\delta_{i}^{D,h}$ for the new variational space. 

\end{enumerate}
The iterative  process stops when the required
precision is reached for the minimized states, i.e.,
when the ML model produces the same lowest energy states as in the previous iteration up to a certain decimal place.
When the model converges, the error  for the low energy states, $|E-\tilde{E}|$, is lower
than the energy difference $|E^{0,0}_A- E^{0,0}_B|$ between competing states $A$ and $B$ near  the phase boundaries.

To obtain a minimal model which includes only the physically
relevant terms, we apply an $L1$ norm regularization of the cost function ~\cite{Tibshirani1996, FujitaH2018}. The details of this procedure are described in the next section. Once the $L1$ regularization eliminates the irrelevant terms in the Hamiltonian, we optimize the model again without the regularization term 
In this way, the powerful technique of $L1$ regularization
outputs a minimal, yet accurate, low energy model Hamiltonian.

\section{Sparse Modeling }
 In setting up the trial Hamiltonian, it is usually not clear how to
determine the number of terms to be included in the initial Hamiltonian.
Too many terms can lead to over-fitting, reducing  the generalization
capability of the acquired model. Furthermore, a $\emph{minimal}$
effective Hamiltonian must include the minimum number of symmetry allowed 
terms necessary to reproduce the phase diagram and the values of relevant physical observables that are predicted by the high-energy model.  
After beginning with an initial Hamiltonian comprising of all the RKKY interaction up to sixth neighbor and all the four-spin interactions up to second neighbor, we implement the $L1$ norm regularization of the cost function to eliminate the least important interactions: 

\begin{equation}
\text{Cost } \!\!\! \left(\left\{ c_{j}\right\} \right) =
\frac{1}{\tilde{N}} \sum_{D,h} \sum_{ i=0}^{\tilde{N}-1} \delta_{i}^{D,h}\left(E_{i}^{D,h}-\tilde{E}_{i}^{D,h}\right)^{2} \!\!\! +\lambda\sum_{j}\vert c_{j}\vert .
\label{eq-L1-CF}
\end{equation}

The added second term in the cost function, with a positive $\lambda $,
penalizes large values of the coupling constants.
We could have also used $L2$ regularization which adds ``squared magnitude'' of the coefficients as penalty term to the loss function instead of the ``absolute value of magnitude'' of coefficients in $L1$ regularization. Both $L1$ and $L2$ regularizations avoid over-fitting, but the key difference is that while $L2$ pushes the coefficients to become small, $L1$ regularization gives sparse estimates (in a high dimensional space it shrinks the less important features' coefficients to zero). Since our main purpose is to obtain the simplest model that reproduces the low-energy physics of the original high-energy model,  $L1$ regularization is the preferred option. The optimal solution is then obtained by
minimizing the new cost function with the gradient descent method
or equivalently with multivariate weighted regression with $L1$ penalty. Once this procedure eliminates the least important interactions, we get an ansatz for the minimal Hamiltonian. The new ansatz is then optimized without the regularization term to get the actual estimation of the  minimal model.

\begin{figure}
\includegraphics[width=\columnwidth]{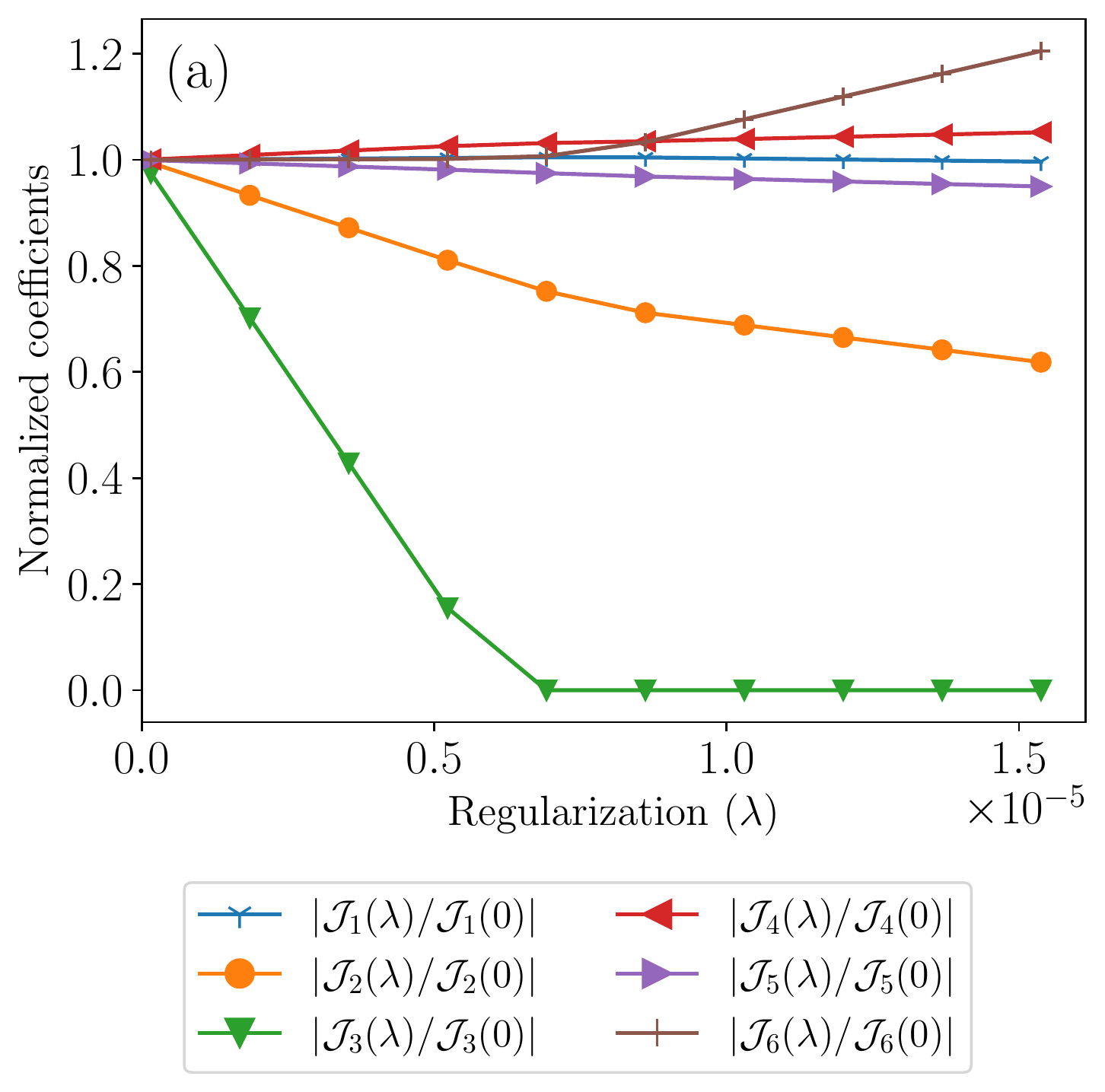}
\includegraphics[width=\columnwidth]{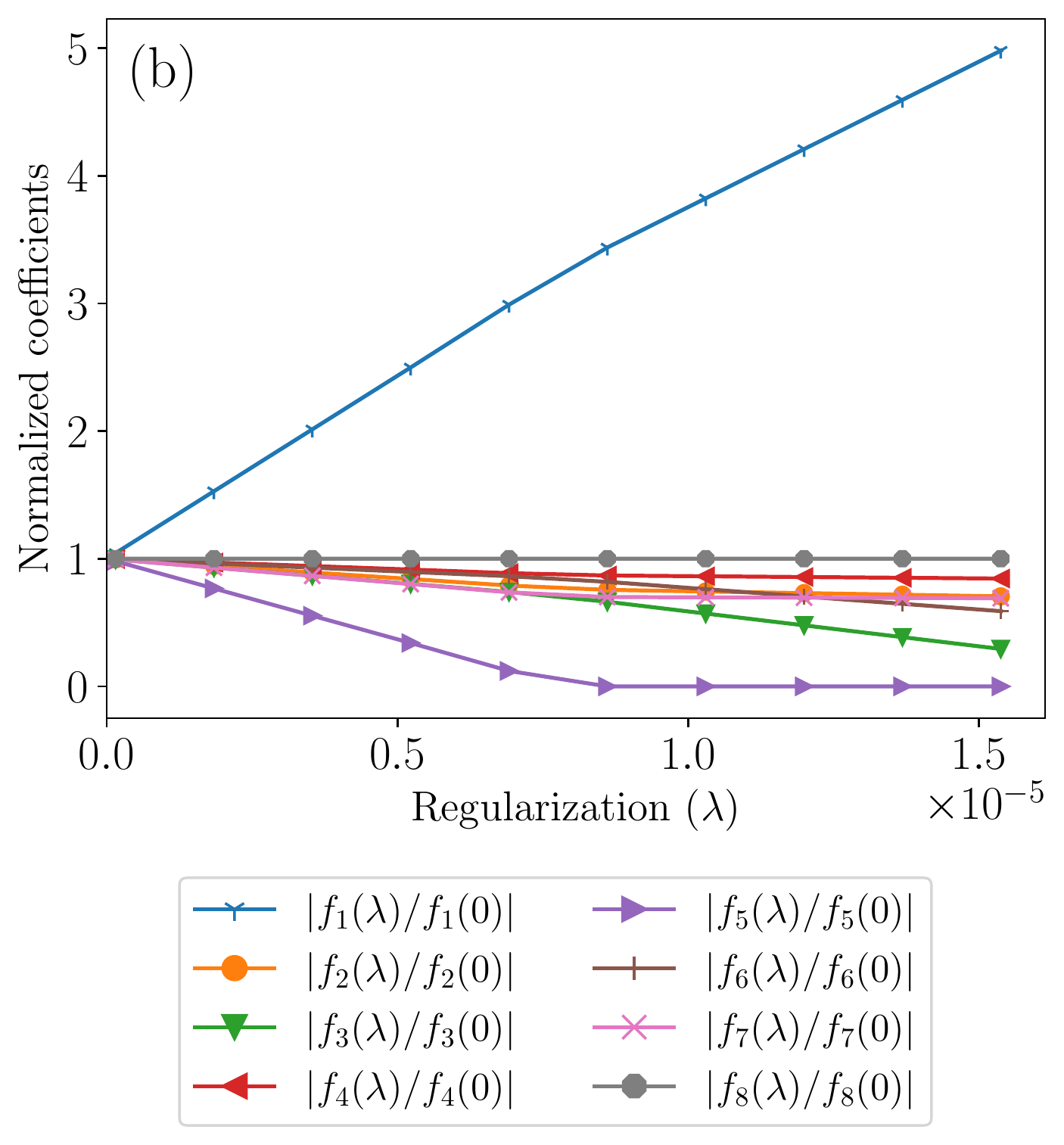}
\caption{ Evolution of parameters as the strength of $L1$ regularization is increased. Each parameter is normalized by dividing its values at various $\lambda $s with its value at $\lambda = 0$. 
}
\label{L1_graph}
\end{figure}

Determining the  range for hyperparameter $\lambda$ to deduce the most important features can be challenging.  We first calculated the weighted mean square error of the full data (weight for each data entry is given by $\delta_i^{D,h}$) and then selected a range of $\lambda$ so that the contribution from the  regularization term was in the range from zero to $ 100 \% $ of the calculated weighted mean square error. By using this range, we avoided the problem of too high regularization penalty. As the value of $\lambda$ is gradually increased, $J_3$ and $f_5$ go to zero  (see Fig.~\ref{L1_graph}). After eliminating these two interactions, we select the final ansatz i.e., the remaining interactions and apply the last step of our algorithm to get the minimal model (Table.~\ref{tab:real space parameters}). In principle, we could have eliminated more interactions by increasing the value of $\lambda$ further, but that would have resulted in a less accurate low energy model.

\begin{table}[tbp!]
\begin{tabular}{rrrrrr} 
\hline  	&	Iteration 1 & Iteration 2 & Iteration 3 & Iteration 4 &  After $L1$
\\  \hline $\mathcal{J}_1$	&	-2.35165 & -2.64772 & -2.64827 & -2.64978 & -2.68884 
\\  $\mathcal{J}_2$	&	-0.39479 & -0.25397 & -0.25489 & -0.25327 & -0.21033 
\\  $\mathcal{J}_3$	&	-0.07018 &  0.01546 &  0.01673 &  0.01721 &  \slashbox{}{}~~~~~ 
\\   $\mathcal{J}_4$	&	0.22517 &  0.25812 &  0.25730 &  0.25776 &  0.26126 
\\   $\mathcal{J}_5$	&	0.08376 &  0.15388 &  0.15375 &  0.15422 &  0.15312 
\\   $\mathcal{J}_6$	&	0.21353 &  0.07107 &  0.07125 &  0.07081 &  0.05770 
\\ \hline $f_1$	&	-0.07113 &  0.01635 &  0.02298 &  0.01984 &  0.02927 
\\  $f_2$	&	-0.75221 & -0.33654 & -0.33809 & -0.33496 & -0.27487 
\\  $f_3$	&	-0.13638 & -0.12970 & -0.13573 & -0.13209 & -0.14971 
\\  $f_4$	&	-0.10768 & -0.09602 & -0.09395 & -0.09633 & -0.08659 
\\   $f_5$	&	0.28331 &  0.04871 &  0.04790 &  0.04722 &  \slashbox{}{}~~~~~  
\\   $f_6$	&	0.12508 &  0.16483 &  0.16582 &  0.16523 &  0.18360 
\\   $f_7$	&	0.84930 &  0.31596 &  0.31372 &  0.31393 &  0.22350 
\\ \hline 
\end{tabular}
\caption{\label{tab:real space parameters}Real space Hamiltonian parameters for Eqs.~\eqref{eq:RKKY_real} and \eqref{eq:4spin_real} at each iteration (units: $10^{-3}t$).}
\end{table}

\begin{table*}
\begin{tabular}{rrrrrrrrrrr} 
\hline  $\tilde{\mathcal{J}}({\bm{Q}})$ & $\tilde{\mathcal{J}}({\bm{0}})$ & $\tilde{g}_{1}$ & $\tilde{g}_{2}$ & $\tilde{g}_{3}$ & $\tilde{g}_{4}$ & $\tilde{g}_{5}$ & $\tilde{g}_{6}$ & $\tilde{g}_{7}$ & $\tilde{g}_{8}$ & $\tilde{g}_{0}$ 
\\  -6.245024 & -6.497512 & -0.020120 & 1.627376 & -0.347720 & 1.592957 & -2.084520 & 1.183781 & -1.380299 & -3.438664 & 0.343609 
\\ \hline 
\end{tabular}
\caption{\label{tab:momentum space parameters}Momentum space Hamiltonian parameters for Eq.~\eqref{eq:model_ham_mom} in units of $10^{-3}t$.} 
\label{TableII}
\end{table*}

\section{Data from High Energy Model - KLM Calculations \label{KLM}}
When the Kondo exchange interaction is comparable to the nearest neighbor
hopping ($J \lesssim t$), the effective spin-spin interactions are orders of magnitude
smaller than the bare Hamiltonian parameters. This leads to large finite
size effects that can not be ignored -- Even for very large system sizes,
the relative stability of two competing states can switch in comparison
to the thermodynamic limit. To avoid these undesirable effects, it is
imperative to work in the thermodynamic limit. Thus, we implemented
a variational approach on a fixed magnetic unit cell~\cite{WangZ2022a}. In the following,
we assume that the magnetic unit cell is spanned by the basis $\left\{ L\bm{a}_{1},L\bm{a}_{2}\right\} $
where $\bm{a}_{1}$ and \textbf{$\bm{a}_{2}$ }are the primitive
vectors of the lattice. We label different sites in the magnetic
unit cell by $\bm{R}$ and the different magnetic unitcells
 by $\tilde{\bm{r}}$. Coordinates of each site
can be expressed as $\bm{r}=\tilde{\bm{r}}+\bm{R}$.
By using the translation symmetry of commensurate states, we can write
the Fourier transform as
\begin{equation}
c_{\tilde{\bm{r}},\bm{R},\sigma}=\sqrt{\frac{L^{2}}{N}}\sum_{\tilde{\bm{k}}}e^{\iu \tilde{\bm{k}}\cdot \tilde{\bm{r}}}c_{\tilde{\bm{k}},\bm{R},\sigma},
\end{equation}
where $\tilde{\bm{k}}$ labels are the allowed momenta in the
reduced Brillouin zone $\mathcal{B}_{\bm{r}}$ and $N$ is the
total number of sites. The KLM Hamiltonian is block-diagonal in momentum
space, $\mathcal{H}=\sum_{\tilde{\bm{k}}}\mathcal{H}_{\tilde{\bm{k}}}$,
where
\begin{align}
\mathcal{H}_{\tilde{\bm{k}}} &=  \sum_{\bm{R}} \Big[-t\sum_{\eta}\sum_{\sigma}c_{\tilde{\bm{k}},\bm{R},\sigma}^{\dagger}c_{\tilde{\bm{k}},\bm{R}+\bm{r}_{\eta},\sigma}  
\nonumber \\ 
&\quad + J\sum_{\alpha\beta}c_{\tilde{\bm{k}},\bm{R},\alpha}^{\dagger}\bm{\sigma}_{\alpha\beta}c_{\tilde{\bm{k}},\bm{R},\beta}\cdot \bm{S}_{\bm{R}}-hS_{\bm{R}}^{z}+D\left(S_{\bm{R}}^{z}\right)^{2}
\Big].
\end{align}
Here, \{$\bm{r}_1$, \ldots, $\bm{r}_6$\} denote the relative position of the nearest neighbor sites. 
Note that Bloch's theorem is   implied here: 
\begin{equation}
c_{\tilde{\bm{k}},\bm{R}+\tilde{\bm{r}},\sigma} \equiv e^{\iu \bm{k} \cdot \tilde{\bm{r}}} c_{\tilde{\bm{k}},\bm{R},\sigma}.
\end{equation}

A $2L^{2}\times2L^{2}$ block
matrix is diagonalized for each $\tilde{\bm{k}}$  to obtain single particle eigenstates. The $T=0$ energy
density is then computed as:
\begin{align}
\frac{E}{N}&=\frac{1}{N}\sum_{\tilde{\bm{k}}}\sum_{n=1}^{2L^{2}}\Theta\left(\mu-\epsilon_{n,\tilde{\bm{k}}}\right)\epsilon_{n,\tilde{\bm{k}}} \nonumber \\
&\quad +\frac{1}{L^{2}}\sum_{\bm{R}}\left[-hS_{\bm{R}}^{z}+D\left(S_{\bm{R}}^{z}\right)^{2}\right],
\end{align}
where $\Theta$ is the step function which selects the energy levels
below the chemical potential $\mu$, and $\epsilon_{n,\tilde{\bm{k}}}$
represent the eigenvalues of the block matrix. In order to accurately identify
the ground state, 
we take the thermodynamic limit by
converting the discrete sum $\frac{1}{N}\sum_{\tilde{\bm{k}}}$
into an integral:
\begin{align}
\frac{E}{N}&= \frac{1}{L^{2}}\int_{\mathcal{B}_{\bm{r}}}\frac{d\tilde{\bm{k}}}{\mathcal{A}_{\mathcal{B}_{\bm{r}}}}\sum_{n=1}^{2L^{2}}\Theta\left(\mu-\epsilon_{n,\tilde{\bm{k}}}\right)\epsilon_{n,\tilde{\bm{k}}}
\nonumber \\
&\quad + \frac{1}{L^{2}}\sum_{\bm{R}}\left[-hS_{\bm{R}}^{z}+D\left(S_{\bm{R}}^{z}\right)^{2}\right],
\end{align}
where $\mathcal{A}_{\mathcal{B}_{\bm{r}}}$ represents the area
of the reduced Brillouin zone. 

In the minimization
process, the chemical potential $\mu$ was determined self consistently
from the filling fraction:
\begin{equation}
n_{c}=\frac{1}{2L^{2}}\int_{\mathcal{B}_{\bm{r}}}\frac{d\tilde{\bm{k}}}{\mathcal{A}_{\mathcal{B}_{\bm{r}}}}\sum_{n=1}^{2L^{2}}\Theta\left(\mu-\epsilon_{n,\tilde{\bm{k}}}\right).
\end{equation}
The various phases of the phase diagram were then obtained by minimizing
the ground state energy over all the possible spin structures for
the fixed magnetic unit cell. For each set of parameters, we typically
performed at least 20 independent minimization runs with different initial
spin configurations  to avoid metastable local minima~\cite{WangZ2022a}.

\begin{figure}
\centering
\includegraphics[width=\columnwidth]{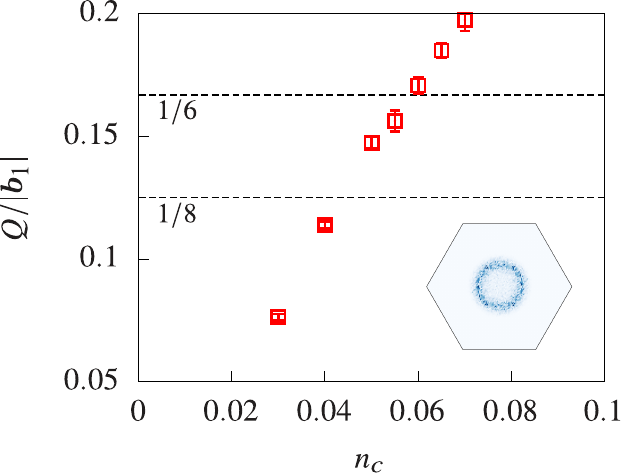}
\caption{Ordering wave number as a function of the filling fraction for the triangular KLM on a $96\times96$ lattice with $h=D=0$, $J/t=0.5$, and $T=10^{-5} J^2/t$, obtained from KPM-SLL simulation. The inset shows a snapshot of the static spin structure factor $\mathcal{S}(\bm{q})$ in the first Brillouin zone with $n_c=0.06$. This figure is reproduced from Ref.~\cite{WangZ2022a}.}
\label{fig:nc}
\end{figure}

The immediate challenge is to find the right size
of the magnetic unit cell, which is a requirement to validate the variational scheme. In the RKKY limit ($J/t\rightarrow0$),
ordering wave vectors $\bm{Q}_{\nu}\,\,\left(\nu=1,2,3\right)$ are located along the high-symmetry $\Gamma$-M directions with  magnitude $\vert\bm{Q}_{\nu}\vert=2k_{F}$~\cite{WangZ2020}. 
However, a finite value of $J/t$ leads to a shift of the wave vectors~\cite{WangZ2016a}. To find the
correct  values,
we simulated the KLM on a $96\times 96$ lattice using stochastic Landau-Lifshitz (SLL) dynamics where KPM was used to obtain the free energy and local forces~\cite{WeisseA2006_RMP,BarrosK2013,WangZ2018}.
Even though a typical finite lattice  is not adequate to accurately compute the relative energies 
of competing states,
it is sufficient to determine the size of the magnetic unit cell.
KPM-SLL represents
a completely unbiased approach to find the period of the optimal ground
state ordering. However, since this method is not effective in handling zero temperature, here we introduced
a very small temperature.
For $J/t=0.5$, temperature $T=10^{-5}J^{2}/t$, $D=0$ and $h=0$, we integrated the dimensionless SLL dynamics
with a unit damping parameter using the Heun-projected scheme for a total of 45000 steps of duration 
$\Delta\tau=\frac{0.5}{\left(J^{2}/t\right)}$.
We used the gradient-based probing method with $S = 256$ colors and $M = 1000$ for
the order of the Chebyshev polynomial expansion~\cite{WangZ2018}. We discarded
the first 30000 steps for equilibration and used the rest 15000 steps
for measurements. Final results were averaged over 6 independent runs
to estimate the error bars. To get a magnetic unitcell of size $6\times6$, the
results yielded a filling fraction $n_{c}\approx 0.0586$ (see Fig.~\ref{fig:nc}).

\section{Phase diagram of the ML model \label{results}}
\begin{figure*}
\includegraphics[width=\textwidth]{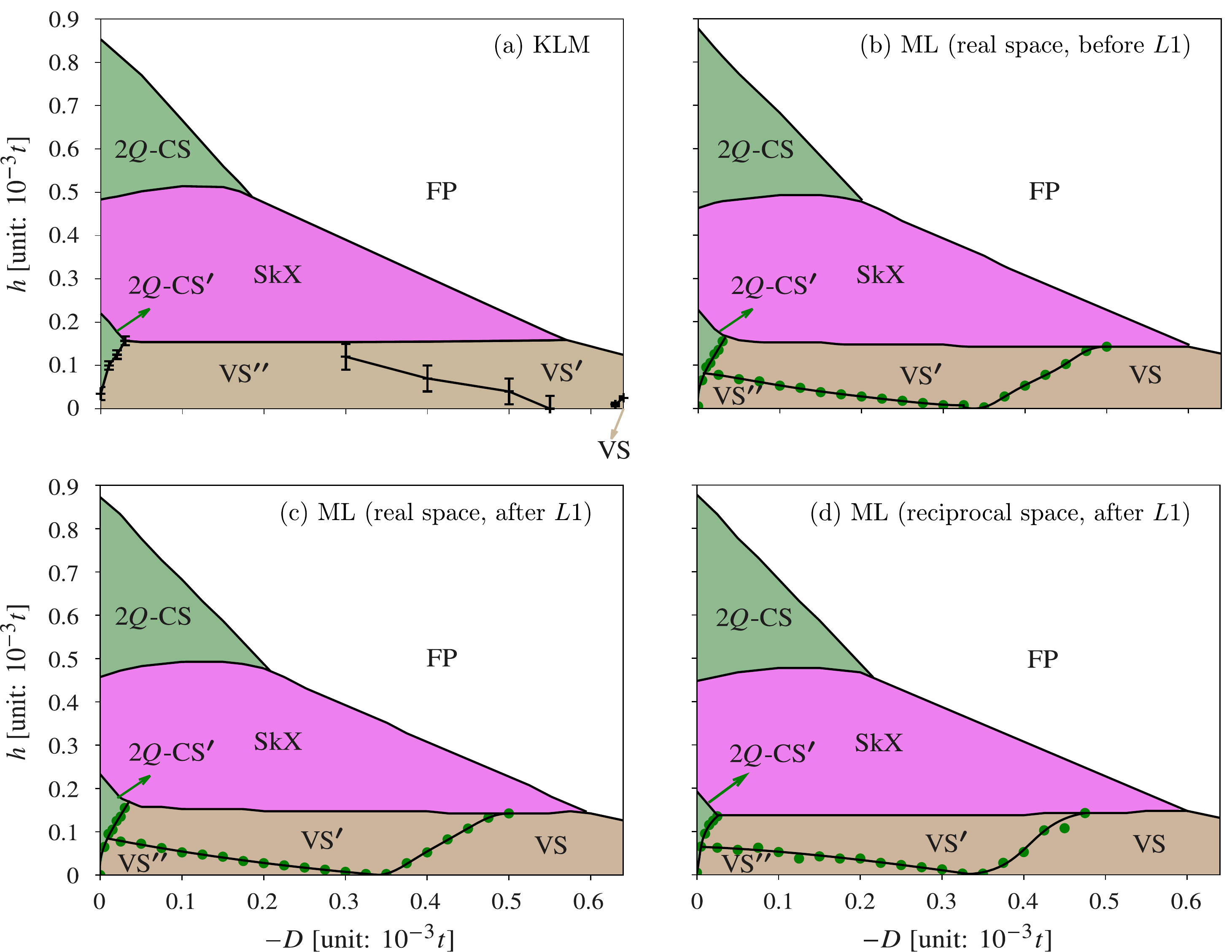}
\caption{$T=0$ phase diagrams of the KLM at $J/t=0.5$ and $n_c=0.0586$. (a) phase diagram obtained via high energy model (KLM) [reproduced from Fig.~2(b) of Ref.~\cite{WangZ2022a}]. The error bars of phase boundaries at low field indicate the limited
numerical accuracy due to quasi-degenerate states. (b) phase diagram of the real space spin model obtained via ML. (c) phase diagram corresponding to real space ML model after $L1$ regularization, and (d) phase diagram corresponding to the minimal momentum space model obtained from $L1$ regularized real space ML model. The phase boundaries at low field in (b-d) are denoted by the filled circles, where the corresponding lines are guides to the eye.}
\label{Phase Diagrams}
\end{figure*}

Our algorithm took four iterations to converge for $J/t = 0.5$. Table~\ref{tab:real space parameters} shows the real space Hamiltonian parameters  for each iteration and the final minimal model. 
Note that all the exchange parameters are about three orders of magnitude smaller than the bare energy scales ($t$, $J$) in the original model, which again explains the difficulty of directly solving the KLM numerically.
The first neighbor RKKY interaction ($\mathcal{J}_1$) is an order of magnitude larger than the next biggest interaction, and the magnitude of four-spin interactions is comparable to the other RKKY interactions, confirming that the KLM parameters are in the intermediate coupling regime.  

To obtain the phase diagram of the ML model, we generated random spin configurations for each $h$ and $D$ and used the gradient descent method to minimize the energy. To find the global minimum for each point, we performed multiple independent runs with different initial random spin configurations. The phase diagrams  obtained from variational calculations  of the original KLM and the ML models are given in Fig.~\ref{Phase Diagrams}. These diagrams include seven different phases: the vertical spiral (VS), vertical
spiral with  in-plane modulation along one direction (VS$^{\prime\prime}$), vertical
spiral with  in-plane modulation (VS$^{\prime}$), 2$\bm{Q}$-conical
spiral (2$\bm{Q}$-CS), 2$\bm{Q}$-conical spiral with unequal
in-plane structure factor intensities(2$\bm{Q}$-CS$^{\prime}$), SkX and
the fully polarized (FP) phase,
where the real space spin configurations of all the phases are given in Refs.~\cite{WangZ2020, WangZ2022a}.
The ML models reproduce the phase boundaries  with remarkable accuracy. Even the simplest minimal model in momentum space (Table~\ref{tab:momentum space parameters}) is able to predict the correct phase boundaries.    We should note, however, that this might not be always the case for the minimal model in momentum space if the ground state magnetic orderings have sizable weight in higher harmonics. 
 For the phase transitions: VS$^{\prime\prime}$ to VS$^\prime$ and VS$^\prime$ to VS, the phase boundaries predicted by the ML models are qualitatively accurate but quantitatively they do not reproduce the KLM phase boundaries. This shortcoming of the  low energy ML model can be attributed to the extremely small energy differences between the three competing phases. Magnitudes of the $\bm{q}$-Fourier components  that differentiate between VS, VS$^{\prime}$ and VS$^{\prime\prime}$ states are orders of magnitude smaller than the magnitude of major $\bm{Q}$-component. The exceeding proximity between the energies of the phases makes it hard to distinguish by a limited number of spin interaction terms in the ML Hamiltonian.

The next question that we can ask is: what can we learn from the effective low-energy model derived with the ML protocol? 
To answer this question, we will consider the problem of understanding the stabilization mechanism of the SkX phase. 
As it has been shown in a recent work~\cite{WangZ2020}, a finite easy-axis anisotropy is required to stabilize the field-induced SkX phase of the  RKKY model. Since the phase diagram of the KLM for $J/t=0.5$ includes a field-induced SkX phase for $D=0$, it is clear that the effective four spin interactions are responsible for its stabilization  in absence of single-ion anisotropy. It is then interesting to inquire about the nature of the four-spin interactions that stabilize the SkX phase. The low-energy effective spin model given in Eq.~\eqref{eq:4spin_g_energy} includes two types of four-spin interactions: those involving the uniform spin component  $\bm{S}_{\bm{0}}$ and those that only involve the finite-$\bm{Q}$ Fourier components $\bm{S}_{\bm{Q}_{\nu}}$ with $\nu=1,2,3$. The field induced SkX phase has been traditionally attributed to 
 the $ \tilde{g}_{3} $ interaction~\cite{Garel82}, which is only present in  three-fold symmetric systems, such as the one under consideration, and that involves the uniform field-induced component $\bm{S}_{\bm{0}}$. The simple reason is that this contribution is finite only for spin configurations that have finite Fourier components $\bm{S}_{\bm{Q}_{1}}$, $\bm{S}_{\bm{Q}_{2}}$ and $\bm{S}_{\bm{Q}_{3}}$. However, the negative sign of $ \tilde{g}_{3} $  that we are obtaining from the ML model (see Table~\ref{TableII}) indicates that this term actually increases the energy of the SkX phase relative to the competing single-$\bm{Q}$ and double-$\bm{Q}$ orderings. The prevalent interactions that assist in the stabilization of SkXs are $ \tilde{g}_{8} $  and $ \tilde{g}_{5} $ -- They have the biggest magnitude in the minimal model and the large negative coefficients correspond to an attractive interaction between pairs of different modes that lowers the energy for triple-$\bm{Q}$ and double-$\bm{Q}$ orderings relative to the single-$\bm{Q}$ vertical spiral ordering. The magnitude of their negative contribution to the SkX energy is higher than that of double-$\bm{Q}$ spiral and thus they help stabilize the SkX phase. We note that the single-$\bm{Q}$ vertical spiral ordering is stable at low enough fields because it is the only phase that has zero magnetization, implying that the weight $|\bm{S}_{\bm{Q}_1}|^2$ is higher than the sum of the three weights $|\bm{S}_{\bm{Q}_1}|^2+ |\bm{S}_{\bm{Q}_2}|^2 + |\bm{S}_{\bm{Q}_3}|^2$ of the SkX and the double-$\bm{Q}$ phases. As we will discuss in the last section, this observation has important consequences for the stabilization of SkX phases in tetragonal materials~\cite{Kurumaji17,WangZ2021}.

\section{Dynamics \label{DYN}}
To further check the validity of the ML models, in this section we compare the underlying spin dynamics to the one obtained from the original KLM.

For concreteness, in this section we choose a magnetic field $h=0.01t$ that is higher than the saturation field, i. e., strong enough to fully polarize the local spins, and   $D=0$. At very low temperature (linear regime), the semi-classical spin wave dispersion can be obtained by solving the Landau-Lifshitz (LL) equations of motion:
\begin{equation}
\frac{d\bm{S}_i}{dt} = \bm{S}_i \times \frac{d E}{d \bm{S}_i},\label{eq:LL}
\end{equation}
where $E$ is the internal energy of the system.

The dynamical spin structure factor can be obtained by Fourier transforming the spin configurations both in space and time:
\begin{equation}
\mathcal{S}^{ab}(\bm{k},\omega) \approx \frac{\omega}{T} \langle S_{\bm{k}}^a(\omega) S_{-{\bm{k}}}^b(-\omega) \rangle,
\end{equation}
where the prefactor $\omega/T \gg 1$ accounts for the classical-quantum correspondence  factor required to  obtain the quantum mechanical result (linear spin wave theory) from the classical   one~\cite{ZhangS2019}, and 
\begin{equation}
\bm{S}_{\bm{k}}(\omega) \equiv \frac{1}{\sqrt{T_S}} \int_0^{T_S} dt e^{\iu \omega t} \frac{1}{\sqrt{N}} \sum_i e^{-\iu \bm{k} \cdot \bm{r}_i} \bm{S}_i(t).\label{eq:Sk_fourier}
\end{equation}

The above Eqs.~\eqref{eq:LL}--\eqref{eq:Sk_fourier} can be applied both to the original KLM and to the effective ML models. The main difference is that the cost of computing the local forces $-dE/d\bm{S}_i$ from the KLM is much higher in comparison with the effective low-energy spin models. In this work, the forces of the KLM are always computed with the KPM~\cite{WeisseA2006_RMP,BarrosK2013,WangZ2018,ChernGW2018}.

To obtain $\mathcal{S}(\bm{k},\omega)$ from the KLM, we initialized the spins from the fully polarized state on a $216\times 216$ triangular lattice. 
We then applied the KPM-SLL method (same as in Sec.~\ref{KLM})  to equilibrate the system at $T=10^{-5}t$ with parameters $n_c=0.0586$, $t=1$, $J=0.5$, $D=0$, and $h=0.01$.  We used the Heun-projected scheme with a unit damping parameter and a total of 14000 steps of duration $\Delta \tau = 2.5 / (J^2/t)$. In addition, we adopted $S=324$ colors  for the gradient-based probing. The order of the Chebyshev expansion was set to $M=2000$. 

\begin{figure}
\includegraphics[width=\columnwidth]{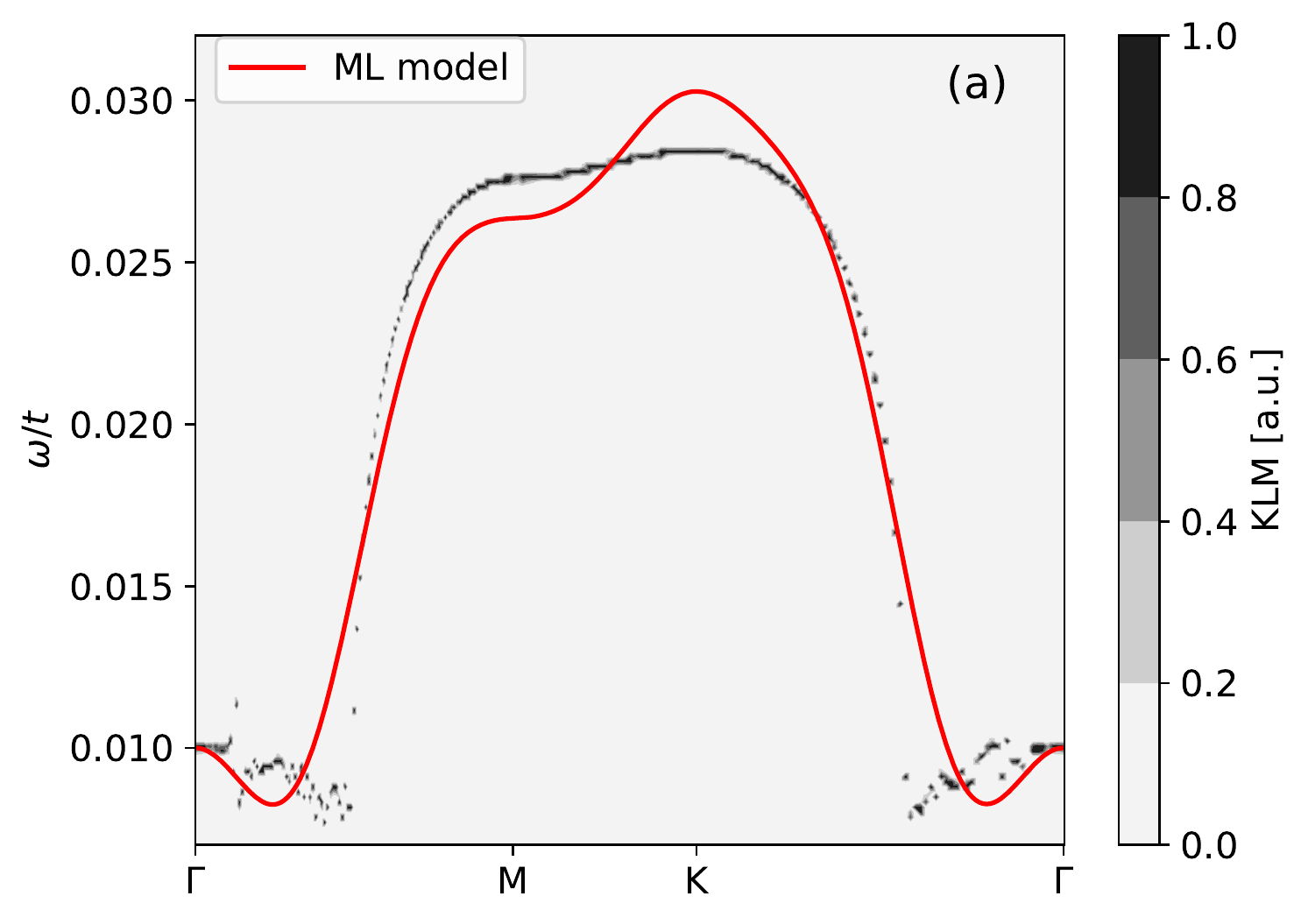}
\includegraphics[width=\columnwidth]{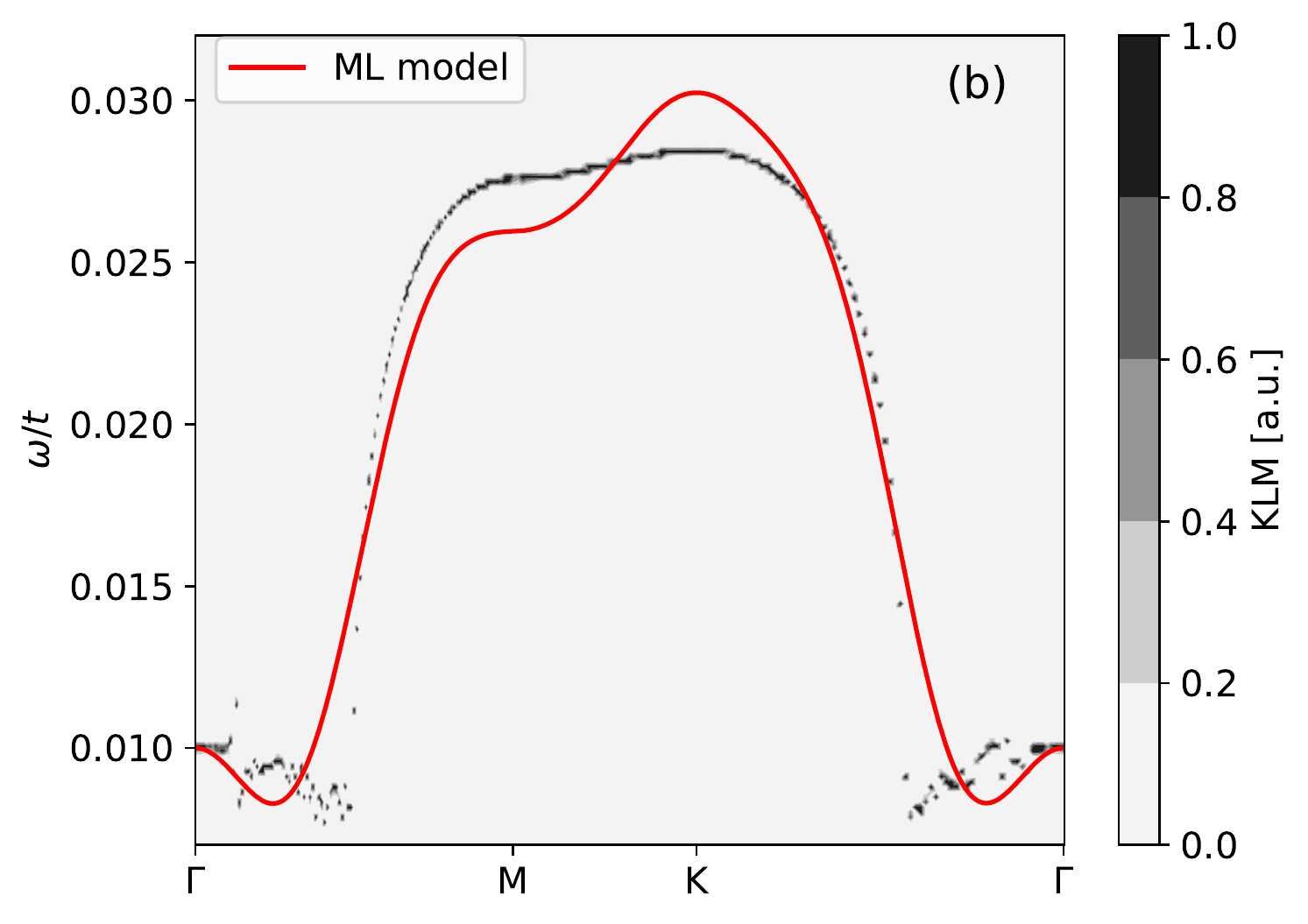}
\caption{Magnon dispersion in the fully polarized phase. (a) Comparison between  linear spin wave results for the ML model before $L1$ regularization and $\mathcal{S}(\bm{k},\omega)$ obtained with the original KLM model. (b) Comparison between the ML model after $L1$ regularization and $\mathcal{S}(\bm{k},\omega)$ obtained with the original KLM model.} 
\label{Magnon Dispersion}
\end{figure}

The spin configuration at the last step was used to seed the LL dynamics~\eqref{eq:LL}, where the damping is set to zero. Once again, the local forces were evaluated using the KPM with $M=2000$ and $S=324$. For convergence, we applied the Heun-projected scheme with a total of 40000 steps of size $\Delta \tau = 0.25/(J^2/t)$ [$T_S = 10^4 /(J^2/t)$ in Eq.~\eqref{eq:Sk_fourier}]. Finally, we used 10 independent runs to compute the average of $\mathcal{S}(\bm{k},\omega)$, which is presented in Fig.~\ref{Magnon Dispersion}.

For the ML models, we can 
calculate the
magnon dispersion analytically in the fully polarized state by implementing the usual Holstein-Primarkoff transformation (linear spin waves):
\begin{equation}
\omega_{\bm{k}}=\Delta E+\sum_{\bm{r}} t_{\bm{r}} \cos\left(\bm{k}\cdot\bm{r}\right),
\end{equation}
where

\begin{equation}
{\begin{aligned}\Delta E & =h+D(1-2S) -6\left(\mathcal{J}_{1}+\mathcal{J}_{2}+\mathcal{J}_{3}+2\mathcal{J}_{4}+\mathcal{J}_{5}+\mathcal{J}_{6}\right)S\\
 &\quad-12\left(f_{1}+f_{3}\right)S^{3} -24\left(f_{2}+f_{4}+2f_{6}\right)S^{3}\\ &\quad-12\left(2f_{5}+f_{7}\right)S^{3},
\end{aligned}
}
\end{equation}

and $t_{\bm{r}}$ depends only on $|\bm{r}|$, with $t_{\bm{r}}=t_1$ for nearest-neighbors, $t_{\bm{r}}=t_2$ second-nearest-neighbors, and so on and so forth:

\begin{align}
t_{1}&=\mathcal{J}_{1}S+\left(2f_{1}+4f_{2}+4f_{6}+4f_{5}+f_{7}\right)S^{3},\\
t_{2}&=\mathcal{J}_{2}S+\left(2f_{3}+4f_{4}+4f_{6}+f_{7}\right)S^{3},\\
t_{\nu}&=\mathcal{J}_{\nu}S\hspace{1em}\{\nu=3,4,5,6\}.
\end{align}

It is quite remarkable that the ML model predicts the magnon dispersion to a reasonable accuracy (Fig.~\ref{Magnon Dispersion}) even though it was never explicitly trained
to reproduce dynamical porperties. The weight factor $\delta^{D,h}$ 
indirectly encodes the information about low-lying excitations in
the model.  The main quantitative discrepancy arises because the magnon dispersion predicted by the ML model an an analytical 
function of $\bm{k}$. In contrast, the magnon dispersion obtained from the KLM seems to have a discontinuous gradient near the band bottom. We  must note, however, that this singular behavior   is originated from  the long-range nature of the effective spin-spin interactions.  Since we are cutting off the bilinear spin interactions beyond sixth neighbors in the ML model,  
the resulting magnon dispersion is necessarily an analytic function of $\bm{k}$. We then expect that the agreement between the magnon dispersions obtained from the high and low-energy models will systematically improve upon including longer range interactions in the effective spin Hamiltonian. Comparisons of the excitation spectrum of both models can then be used as a criterion to fix the cut-off length scale of the  spin-spin interactions included in the low-energy model.

\section{Conclusions and Outlook \label{S&O}}

Our work unravels a novel pathway to derive spin Hamiltonians when  conventional methods, like perturbation theory, fail. One such example is a triangular KLM which gives rise to effective four-spin interactions that are non-analytic functions of $J/t$. We demonstrate that the effective low-energy model derived with  the ML assisted protocol accurately predicts the main phase bpundaries of the phase diagram obtained by directly simulating the KLM. While the  KLM simulations are numerically expensive, simulations of the effective spin model are roughly two orders of magnitude faster for the magnetic unit cell of 36 spins that we used in this work. 

The simplicity of the effective low energy model provides new insight into the stabilization mechanism of the different field-induced multi-$\bm{Q}$ magnetic orderings.
Previous studies of the RKKY model showed that a finite easy-axis anisotropy is required to  stabilize  a magnetic field-induced triple-$\bm{Q}$ SkX phase ~\cite{WangZ2020}. In contrast, as shown in Fig.~\ref{Phase Diagrams}(a), the quantum phase diagram of the KLM includes a field induced SkX phase even in absence of  easy-axis anisotropy~\cite{WangZ2022a}. The source of stabilization of SkX was speculated to be four-spin interactions mediated by itinerant electrons, which are expected to become significant when the ratio $J/t$ is not much smaller than one. The procedure that is described in this work allows us to quantify the magnitude of these four-spin interactions without introducing any bias, other than their limited range, in the types of interactions that can emerge at low energies. In other words, the above-described protocol allows us to include all the symmetry allowed four-spin interactions within a certain range. The good agreement between the phase diagrams of the original KLM and the effective low-energy model suggests that the field-induced SkX phase is indeed stabilized by effective four-spin interactions that turn out to be attractive between pairs of modes  with different 
wave vector $\bm{Q}_{\nu}$. This attractive channel, which was conjectured in previous studies of multi-$\bm{Q}$ orderings in metallic systems~\cite{MartinI2008,AkagiY2012,WangZ2022a}, is indeed expected to be present whenever the  ordering wave vectors $\bm{Q}_{\nu}$ connect different regions of the Fermi surface~\cite{WangZ2022a}. On a more intuitive level, the multi-$\bm{Q}$ ordering gaps out a bigger region of the Fermi surface in comparison to the single-$\bm{Q}$ phase. Moreover, the protocol presented in this work can also be used as a tool to compute the phase diagram of the original KLM, and the acceleration factor is of the order of one hundred. The most  expensive part of the protocol is the generation of the training data set that is much smaller than the set required to build the phase diagram of the original KLM.

As we discussed in Sec.~\ref{results}, the low-energy spin model derived with the ML assisted protocol provides new insights into the stabilization mechanism of the SkX phase in KLMs. After recognizing that the stabilization arises from an attractive interaction between pairs of different modes 
\{$\bm{Q}_{\nu}$, $\bm{Q}_{\nu'}$\} represented by the $ \tilde{g}_{8} $  and $ \tilde{g}_{5} $ terms of Eq.~\eqref{eq:4spin_g_energy}, and that the $ \tilde{g}_{3} $ contribution is actually increasing the energy of the SkX relative to the competing orderings, we can infer that  field-induced square SkX phases should also arise in KLMs with tetragonal symmetry. The key observation is that the $ \tilde{g}_{3} $ term is no longer relevant because the square SkX is a double-$\bm{Q}$ ($\bm{Q}_1$ and $\bm{Q}_2$) ordering with $\bm{Q}_1 \cdot \bm{Q}_2=\bm{0}$. This means that the square SkX phase will still benefit from the attractive interaction between the pair of different modes, while the penalization from a $ \tilde{g}_{3}$-like term with $\bm{Q}_3 =-\bm{Q}_1 - \bm{Q}_2$ will be significantly smaller because  $\bm{Q}_3$ is no longer a fundamental wave-vector, but a second harmonic ($|\bm{S}_{\bm{Q}_3}| \ll |\bm{S}_{\bm{Q}_1}|, |\bm{S}_{\bm{Q}_2}|$). 
The only remaining obstacle for the stabilization of  the square SkX is the RKKY energy cost of the second harmonic component $|\bm{S}_{\bm{Q}_1+\bm{Q}_2}|$.
In the perturbative regime, $|J| \ll |t|$, this energy cost of order $ J^2 |\bm{S}_{\bm{Q}_1+\bm{Q}_2}|^2   (\chi_{\bm{Q}_1}-\chi_{\bm{Q}_1+\bm{Q}_2})$ can be reduced by choosing tetragonal materials/models with square-like Fermi surfaces. 
In other words, square SkXs should emerge for  $ J^2 |\bm{S}_{\bm{Q}_1+\bm{Q}_2}|^2   (\chi_{\bm{Q}_1}-\chi_{\bm{Q}_1+\bm{Q}_2}) \lesssim \tilde{g}_{8} |\bm{S}_{\bm{Q}_1}|^2 |\bm{S}_{\bm{Q}_2}|^2$.

 Finally, it is interesting to note that the $ \tilde{g}_{7}$ biquadratic interaction, which has been adopted in previous works as the only 4-spin interaction generated by the KLM ~\cite{HayamiS2017,HayamiS2021_review}, is revealed by our ML protocol to be a subdominant term (see Table~\ref{tab:momentum space parameters}). Since this biquadratic interaction was selected based on a trend  observed in the divergent terms of the perturbative expansion in $J/t$, we conclude that such procedure is not reliable to quantify the relative strength of the different symmetry allowed four-spin interactions.

These simple examples illustrate how ML-assisted protocols can be used to extract guiding principles. While it is natural to assume that the $ \tilde{g}_{3} $ term is responsible for the stabilization of SkXs in hexagonal lattices~\cite{Garel82}, this intuitive argument turned out be incorrect. The ML-assisted protocol not only allows us to correct this assumption, but also gives us enough information to  infer a new guiding principle. In other words, besides providing an efficient tool to accelerate the computation of phase diagrams, the ML approach is also an efficient learning tool that can be used to understand mechanisms and accelerate discovery. 

\section*{ACKNOWLEDGEMENTS}
We thank Kipton Barros for useful discussions. This work was supported by the U.S. Department of Energy, Office of Science, Basic Energy Sciences, under Award No.~DE-SC0022311.

\FloatBarrier

\bibliography{ref}




\appendix

\section{Hamiltonian in momentum space}\label{sec:g_expression}
The mathematical expression for the four-spin interactions' contribution
to the Hamiltonian is written as:
\begin{equation}
\mathcal{H}_{4\text{-spin}}=\sum_{\langle \bm{r}_1,\bm{r}_2,\bm{r}_3,\bm{r}_4 \rangle}f(\bm{r}_1,\bm{r}_2,\bm{r}_3,\bm{r}_4) \left(\bm{S}_{\bm{r}_1}\cdot \bm{S}_{\bm{r}_2}\right)\left(\bm{S}_{\bm{r}_3}\cdot \bm{S}_{\bm{r}_4}\right).
\end{equation}
In momentum space, it becomes:
\begin{equation}
\mathcal{H}_{4\text{-spin}}=\frac{1}{N}\sum_{\bm{k}_{2},\bm{k}_{3},\bm{k}_{4}}g\left(\bm{k}_{2},\bm{k}_{3},\bm{k}_{4}\right)\left(\bm{S}_{-\bm{k}_{2}-\bm{k}_{3}-\bm{k}_{4}}\cdot\bm{S}_{\bm{k}_{2}}\right)\left(\bm{S}_{\bm{k}_{3}}\cdot\bm{S}_{\bm{k}_{4}}\right),
\end{equation}
where
\begin{align}
g\left(\bm{k}_{2},\bm{k}_{3},\bm{k}_{4}\right) & = \sum_{\langle \bm{r}_{21},\bm{r}_{31},\bm{r}_{41}\rangle }f(\bm{0},\bm{r}_{21},\bm{r}_{31},\bm{r}_{41}) \nonumber \\
 & \quad \cdot e^{-\iu \left( \bm{k}_{2}\cdot \bm{r}_{21} + \bm{k}_{3}\cdot \bm{r}_{31} + \bm{k}_{4}\cdot \bm{r}_{41} \right)},   
\end{align}
where $\bm{r}_{ij}\equiv \bm{r}_i - \bm{r}_j$ is the relative position of two sites in the four-spin interactions.

If we limit the possible $\bm{k}$-values to set of ordering
wave vectors $\left(\bm{Q}_{1},\bm{Q}_{2},\bm{Q}_{3},\bm{0}\right)$,
we get only a few possible combinations:
\begin{subequations}
\begin{align}
\tilde{g}_{0}(\bm{0},\bm{0},\bm{0})&=g(\bm{0},\bm{0},\bm{0}),\\
\tilde{g}_{1}\left(\bm{Q}_{i},\bm{0},\bm{0}\right) & =g\left(\bm{Q}_{i},\bm{0},\bm{0}\right)+g\left(\bar{\bm{Q}}_{i},\bm{0},\bm{0}\right)\nonumber \\
 &\quad  +g\left(\bm{0},\bm{Q}_{i},\bar{\bm{Q}}_{i}\right)+g\left(\bm{0},\bar{\bm{Q}}_{i},\bm{Q}_{i}\right), \\
\tilde{g}_{2}\left(\bm{Q}_{i},\bar{\bm{Q}}_{i},\bm{0}\right) & =g\left(\bm{0},\bm{0},\bm{Q}_{i}\right)+g\left(\bar{\bm{Q}}_{i},\bm{0},\bm{Q}_{i}\right)\nonumber \\
 & \quad +g\left(\bm{0},\bm{0},\bar{\bm{Q}}_{i}\right)+g\left(\bar{\bm{Q}}_{i},\bm{Q}_{i},\bm{0}\right)\nonumber \\
 & \quad +g\left(\bm{0},\bm{Q}_{i},\bm{0}\right)+g\left(\bm{Q}_{i},\bar{\bm{Q}}_{i},\bm{0}\right)\nonumber \\
 & \quad +g\left(\bm{Q}_{i},\bm{0},\bar{\bm{Q}}_{i}\right)+g\left(\bm{0},\bar{\bm{Q}}_{i},\bm{0}\right),\\
\tilde{g}_{3}\left(\bm{Q}_{1},\bm{Q}_{2},\bm{Q}_{3}\right) & =g\left(\bm{Q}_{3},\bm{Q}_{1},\bm{0}\right)+g\left(\bm{0},\bm{Q}_{2},\bm{Q}_{3}\right)\nonumber \\
 & \quad +g\left(\bm{Q}_{2},\bm{Q}_{1},\bm{0}\right)+g\left(\bm{Q}_{1},\bm{Q}_{2},\bm{Q}_{3}\right)\nonumber \\
 & \quad +g\left(\bm{Q}_{2},\bm{0},\bm{Q}_{1}\right)+g\left(\bm{0},\bm{Q}_{3},\bm{Q}_{2}\right)\nonumber \\
 & \quad +g\left(\bm{Q}_{1},\bm{Q}_{3},\bm{Q}_{2}\right)+g\left(\bm{Q}_{3},\bm{0},\bm{Q}_{1}\right), \\
\tilde{g}_{4}\left(\bar{\bm{Q}}_{j},\bar{\bm{Q}}_{i},\bm{Q}_{j}\right) & =g\left(\bar{\bm{Q}}_{i},\bm{Q}_{i},\bar{\bm{Q}}_{j}\right)+g\left(\bm{Q}_{j},\bm{Q}_{i},\bar{\bm{Q}}_{j}\right)\nonumber \\
 &\quad  +g\left(\bm{Q}_{i},\bar{\bm{Q}}_{i},\bm{Q}_{j}\right)+g\left(\bar{\bm{Q}}_{j},\bar{\bm{Q}}_{i},\bm{Q}_{j}\right)\nonumber \\
 & \quad +g\left(\bar{\bm{Q}}_{j},\bm{Q}_{j},\bar{\bm{Q}}_{i}\right)+g\left(\bm{Q}_{i},\bm{Q}_{j},\bar{\bm{Q}}_{i}\right)\nonumber \\
 & \quad +g\left(\bar{\bm{Q}}_{i},\bar{\bm{Q}}_{j},\bm{Q}_{i}\right)+g\left(\bm{Q}_{j},\bar{\bm{Q}}_{j},\bm{Q}_{i}\right),\\
\tilde{g}_{5}\left(\bm{Q}_{j},\bar{\bm{Q}}_{i},\bar{\bm{Q}}_{j}\right) & =g\left(\bm{Q}_{j},\bar{\bm{Q}}_{i},\bar{\bm{Q}}_{j}\right)+g\left(\bm{Q}_{j},\bar{\bm{Q}}_{j},\bar{\bm{Q}}_{i}\right)\nonumber \\
 & \quad +g\left(\bar{\bm{Q}}_{j},\bm{Q}_{j},\bm{Q}_{i}\right)+g\left(\bar{\bm{Q}}_{j},\bm{Q}_{i},\bar{\bm{Q}}_{j}\right)\nonumber \\
 & \quad +g\left(\bar{\bm{Q}}_{i},\bm{Q}_{i},\bm{Q}_{j}\right)+g\left(\bm{Q}_{i},\bar{\bm{Q}}_{j},\bar{\bm{Q}}_{i}\right)\nonumber \\
 & \quad +g\left(\bar{\bm{Q}}_{i},\bm{Q}_{j},\bm{Q}_{i}\right)+g\left(\bm{Q}_{i},\bar{\bm{Q}}_{i},\bar{\bm{Q}}_{j}\right),\\
\tilde{g}_{6}\left(\bm{Q}_{i},\bar{\bm{Q}}_{i},\bar{\bm{Q}}_{i}\right) &=g\left(\bar{\bm{Q}}_{i},\bm{Q}_{i},\bm{Q}_{i}\right)+g\left(\bm{Q}_{i},\bar{\bm{Q}}_{i},\bar{\bm{Q}}_{i}\right),\\
\tilde{g}_{7}\left(\bar{\bm{Q}}_{i},\bm{Q}_{i},\bar{\bm{Q}}_{i}\right) & =g\left(\bm{Q}_{i},\bm{Q}_{i},\bar{\bm{Q}}_{i}\right)+g\left(\bar{\bm{Q}}_{i},\bm{Q}_{i},\bar{\bm{Q}}_{i}\right)\nonumber \\
 & \quad +g\left(\bar{\bm{Q}}_{i},\bar{\bm{Q}}_{i},\bm{Q}_{i}\right)+g\left(\bm{Q}_{i},\bar{\bm{Q}}_{i},\bm{Q}_{i}\right),\\
\tilde{g}_{8}\left(\bm{Q}_{i},\bm{Q}_{j},\bar{\bm{Q}}_{j}\right) & =g\left(\bm{Q}_{i},\bm{Q}_{j},\bar{\bm{Q}}_{j}\right)+g\left(\bar{\bm{Q}}_{i},\bm{Q}_{j},\bar{\bm{Q}}_{j}\right)\nonumber \\
 & \quad +g\left(\bm{Q}_{i},\bar{\bm{Q}}_{j},\bm{Q}_{j}\right)+g\left(\bar{\bm{Q}}_{i},\bar{\bm{Q}}_{j},\bm{Q}_{j}\right)\nonumber \\
 & \quad +g\left(\bm{Q}_{j},\bm{Q}_{i},\bar{\bm{Q}}_{i}\right)+g\left(\bar{\bm{Q}}_{j},\bm{Q}_{i},\bar{\bm{Q}}_{i}\right)\nonumber \\
 & \quad +g\left(\bm{Q}_{j},\bar{\bm{Q}}_{i},\bm{Q}_{i}\right)+g\left(\bar{\bm{Q}}_{j},\bar{\bm{Q}}_{i},\bm{Q}_{i}\right),
\end{align}
\end{subequations}
where we have denoted $\bar{\bm{Q}}_i = -\bm{Q}_i$.
Using these expression, the energy contribution of the four-spin interactions is written in Eq.~\eqref{eq:4spin_g_energy}.

Similarly, the RKKY contribution to
the Hamiltonian is:
\begin{equation}
\mathcal{H}_{\rm RKKY}=\frac{1}{2}\sum_{\bm{r}\neq \bm{r}^\prime}{\cal J}(\bm{r}-\bm{r}^\prime)\bm{S}_{\bm{r}}\cdot\bm{S}_{\bm{r}^\prime}.
\end{equation}
In momentum space:
\begin{equation}
\mathcal{H}_{\rm RKKY}=\sum_{\bm{k}} \tilde{\mathcal{J}}(\bm{k})\bm{S}_{\bm{k}}\cdot\bm{S}_{-\bm{k}},
\end{equation}
where
\begin{equation}
\tilde{\mathcal{J}}(\bm{k})=\frac{1}{2} \sum_{\bm{r}}{\cal J}(\bm{r})e^{\dot{\iota}\bm{k}\cdot \bm{r}}.
\end{equation}
If we limit the possible $\bm{k}$-values to the set of ordering
wave vectors $\left(\bm{Q}_{1},\bm{Q}_{2},\bm{Q}_{3},\bm{0}\right)$,
we can write the RKKY energy contribution as:
\begin{equation}
E_\text{RKKY}=\sum_{i=1}^{3}2\tilde{\mathcal{J}}(\bm{Q}_{i})\bm{S}_{\bm{Q}_i}\cdot\bm{S}_{-\bm{Q}_i}+\tilde{\mathcal{J}}(\bm{0})\bm{S}_{\bm{0}}\cdot\bm{S}_{\bm{0}}.
\end{equation}


\end{document}